\documentclass[a4paper,11pt]{article}
\pdfoutput=1
\usepackage{jcappub}
\usepackage{graphicx,caption}
\usepackage{amsmath}
\usepackage{bm}
\usepackage{mathrsfs}
\usepackage{latexsym}
\usepackage{float}
\usepackage{color}
\usepackage[normalem]{ulem}
\usepackage{subcaption}
\usepackage{dcolumn}
\usepackage{multirow}
\usepackage[usenames,dvipsnames]{xcolor}
\usepackage{slashed}
\usepackage{booktabs}
\usepackage{multicol}
 
\newcommand{\SRM}[1]{#1}
\newcommand{\HCD}[1]{#1}
\title{The Impact of Anisotropy on Neutron Star Properties: Insights from $I$-$f$-$C$ Universal Relations}
\author[a]{Sailesh Ranjan Mohanty,}
\author[a]{Sayantan Ghosh,}
\author[a]{Pinku Routaray,}
\author[b]{H.C. Das,}
\author[a,1]{Bharat Kumar \note{Corresponding author.}}
\affiliation[a]{Department of Physics \& Astronomy, National Institute of Technology, Rourkela 769008, India;}
\affiliation[b]{INFN Sezione di Catania, Dipartimento di Fisica,
Via S. Sofia 64, 95123 Catania, Italy.}
\emailAdd{kumarbh@nitrkl.ac.in}
\abstract{
Anisotropy in pressure within a star emerges from exotic internal processes. In this study, we incorporate pressure anisotropy using the Quasi-Local model. Macroscopic properties, including mass ($M$), radius ($R$), compactness ($C$), dimensionless tidal deformability ($\Lambda$), the moment of inertia ($I$), and oscillation frequency ($f$), are explored for the anisotropic neutron star. Magnitudes of these properties are notably influenced by anisotropy degree. Universal $I$-$f$-$C$ relations for anisotropic stars are explored in this study. The analysis encompasses various EOS types, spanning from relativistic to non-relativistic regimes. Results show the relation becomes robust for positive anisotropy, weakening with negative anisotropy. \SRM{The distribution of $f$-mode across $M$-$R$ parameter space as obtained with the help of $C$ -$f$ relation was analyzed for different anisotropic cases.} Using tidal deformability data from GW170817 and GW190814 events, a theoretical limit for canonical $f$-mode frequency is established for isotropic and anisotropic neutron stars. For isotropic case, canonical $f$-mode frequency for GW170817 event is $f_{1.4}= 2.606 ^{+0.457} _ {-0.484}\ \mathrm{kHz}$; for GW190814 event, it is $f_{1.4}= 2.097 ^{+0.124} _ {-0.149}\ \mathrm{kHz}$. These relationships can serve as reliable tools for constraining nuclear matter EOS when relevant observables are measured.}
\begin{document}
\maketitle
\flushbottom
\section{Introduction}
The detection of gravitational waves (GWs) is a crucial objective in astrophysics today, with significant efforts being devoted to this problem globally. Several ground-based experiments and space missions have already been devised and are poised to yield significant discoveries in the near future \cite{Abbott_2009, Harry_2010, Acernese_2006, Accadia_2011, Antonucci_2011}. Furthermore, ongoing efforts are underway to develop a third-generation GW telescope, such as the Einstein Telescope \cite{Punturo_2010} and Cosmic Explorer \cite{galaxies10040090}, which promise even higher levels of sensitivity. The reason why the detection of GWs is so challenging is that they are incredibly weak, necessitating detectors with very high sensitivity, as well as the precise waveform of the signal emitted from astrophysical objects.

Neutron stars (NSs) exhibit oscillations that are regarded as potential sources of GWs, manifesting in various forms including radial \cite{Chandrasekhar_1964, Chanmugam_1977, Kokkotas_2001, Pinku-PRD_2023, sen2022radial,Pinku_mnras_2023} and non-radial \cite{mcdermott_1988, Kunjipurayil_2022, Harish-fmode_2022,Pinku_jcap_2023} modes. When a NS experiences external or internal disturbances, it emits GWs through different oscillation modes known as quasi-normal modes (QNMs), each characterized by the restoring force that brings them back to their equilibrium state. Notable QNMs include the fundamental mode ($f$-mode) \cite{Zhao_2022, PhysRevC.103.035810, Pradhan_2022}, pressure mode ($p$-mode) \cite{PhysRevD.104.123002, Kunjipurayil_2022}, gravity mode ($g$-mode) \cite{Finn_1987, Reisenegger_1992, Zhao_gmode_2022, Lozano_2022, Constantinou_2021, Jaikumar_2021, Wei_2020, Tran_2022}, rotational mode ($r$-mode) \cite{Haskell_2014, Haskell_2015, Jyothilakshmi_2022, Lin_2004, Rezzolla_2002, Michael_2017}, space-time mode ($w$-mode) \cite{Benhar_1999, Bandyopadhyay_2012}, and other modes \cite{Kokkotas1999, Sotani_2011, Flores_2014, Sandoval_2018}. The frequencies of these oscillations are directly linked to the internal structure and composition of the stars \cite{Tianqi_2022}. Theoretical studies indicate that the $f$-mode possesses the highest likelihood of being detected initially, with approximately 10\% of the gravitational radiation attributed to $f$-mode oscillations for $l=2$ \cite{Shibagaki_2020}.

Different modes of oscillation exhibit distinct behaviors depending on the type of star, offering valuable insights. For instance, the signature of the hadron-quark phase transition can be inferred through the observation of both $f$ and $g$-modes in hybrid stars \cite{Sotani_2011}. A study by Flores and Lugones \cite{Flores_2014} suggests that compact objects emitting GWs within the frequency range of $0-1$ kHz may be hybrid stars, while frequencies exceeding 7 kHz could indicate strange stars. However, the nature of compact objects and their GW emissions still pose challenges due to the limitations of our terrestrial detectors, which are unable to detect certain frequency ranges. Nonetheless, constraints on the frequency of various modes can be established by establishing relationships between the frequency and specific properties of NSs. Several approaches have been proposed to link the $f$-mode frequency with various NS properties, including compactness \cite{Andersson_1998}, the moment of inertia \cite{Lau_2010}, and tidal deformability \cite{Chan_2014, Sotani_2021, Bikram_2023}. In this work, we aim to derive such relationships in a model-independent manner, known as universal relations (URs). These URs provide valuable insights into the properties of NSs and their associated oscillation modes.

In literature, there are several URs have already been established between different properties of the NS \cite{Kent_yagi_2013, Kent_yagi_2015, Breu_2016, Rezzolla_2018, Riahi_2019, Gupta_2018, Jiang_2020,universe7040111, Chakrabarti_2014, Haskell_2013, Bandyopadhyay2018}. However, the focus of this study primarily lies on the $I-f-C$ relations specifically for anisotropic NSs. Previous URs proposed thus far have mainly been formulated for isotropic NSs, assuming a matter-energy distribution characterized by an isotropic perfect fluid. However, at extremely high-density regions, the presence of nuclear matter can induce deviations between the tangential and radial components of pressure, leading to the emergence of an anisotropic fluid. To obtain more accurate results, we aim to include the effects of anisotropy in our analysis. Anisotropy in NSs can arise from various factors, such as the influence of a high magnetic field \cite{S.S.Yaza, C.Y.Cardall_2001, K.Ioka_2004, R.Ciolfi_L.Rezzolla, R.Ciolfi_V.Ferrari, Frieben, A.G.Pili,N.Bucciantini}, pion condensation \cite{R.F.Sawyer}, phase transitions \cite{B.Carter}, relativistic nuclear interactions \cite{V.Canuto, M.Ruderman}, core crystallization \cite{S.Nelmes}, and the presence of superfluid cores \cite{W.A.Kippenhahn_Rudolf, Glendenning:1997wn, H.Heiselberg}, among others. Several models have been proposed to incorporate anisotropy within NSs, such as the Bowers-Liang model \cite{BL-Model}, Horvat et al. model \cite{Horvat_2011}, Cosenza et al. model \cite{Cosenza}, and others. In this study, we will primarily focus on the Quasi-Local (QL) model, as proposed by Horvat et al. \cite{Horvat_2011}, to describe anisotropy within NSs. More details about the QL-model will be elaborated in Section \ref{sec:anisotropi_model}.

The presence of pressure anisotropy within NSs has a significant impact on various macroscopic properties, including the mass-radius relation, compactness, surface redshift, moment of inertia, tidal deformability, and non-radial oscillations \cite{Silva_2015, Hillebrandt, Doneva_2012, Bayin, Roupas2021, Deb_2021, Estevez-Delgado2018, Pattersons2021, Rizaldy, Rahmansyah2020, Rahmansyah2021, Herrera2008, Herrera2013, Bhaskar_Biswas, S.Das2021, Roupas2020, Sulaksono2015, Setiawan2019, Silva_2015}. The specific impact on these quantities varies depending on the degree of anisotropy and the choice of the model employed. Biswas and Bose constrained the degree of anisotropy within stars utilizing tidal deformability data from the GW170817 event \cite{Bhaskar_Biswas}. More recently, the $I-$Love$-C$ universal relation for anisotropic NSs has been proposed. By incorporating observational data from GW170817, constraints have been placed on the moment of inertia and radius of canonical anisotropic stars for different degrees of anisotropy \cite{HC_I_LOVE_C}. 

In this study, we introduce the URs for anisotropic NSs for the first time, utilizing a range of models describing unified EOSs such as RMF models (for npe$\mu$, hyperonic npe$\mu$Y, and strange npe$\mu$Ys matter), density-dependent RMF models, and Skyrme-Hartree-Fock (SHF) models \cite{Fortin_2016, Bharat_and_Landry, Landry_2018, Athul_2022, Tuhin_2018, Alam_2016, Vishal_2022_Crustal, Vishal_2022_Pasta}. The unified EOSs closely reproduce the properties of finite nuclei, nuclear matter, and NSs and support $\geq2.0 M_{\odot}$ star. A total of 60 EOSs (35 from RMF models and 25 from SHF models) are considered in this paper. By employing these EOSs, we calculate various macroscopic properties of NSs with varying degrees of anisotropy using the QL-model. The primary objective of this work is to constrain the $f$-mode frequency of anisotropic stars by leveraging recent observational data. Detailed formalism, results, and discussions are presented in the subsequent sections, shedding light on the intricate relationship between anisotropy, macroscopic properties, and the $f$-mode frequency in NSs. 
\section{Theoretical Framework}
\subsection{Stellar Structure Equations}
The stress-energy tensor corresponds to the anisotropic distribution of matter-energy as follows \cite{Estevez-Delgado2018}
\begin{equation}
\label{eq:energy_mom}
T_{\mu v}=\left(\mathcal{E}+P_t\right)u_\mu u_v+P_t g_{\mu v}-\sigma k_\mu k_v \, ,
\end{equation}
where $u^\mu$ represents the four-velocity of the fluid, and $k^\mu$ is a unit space-like four-vector. In addition, $\mathcal{E}$ denotes the energy density, and $\sigma$ represents the anisotropic pressure ($\sigma = P_t - P_r$), where $P_t$ and $P_r$ are the tangential and radial pressure respectively. The four-vectors $u^\mu$ and $k^\mu$ must fulfill the following conditions $u_\mu u^\mu=-1, \quad k_\mu k^\mu=1, \quad u_\mu k^\mu=0$.

The stellar structure can be obtained by solving the Tolman-Oppenheimer-Volkoff (TOV) equations for the anisotropic matter-energy distribution as follows \cite{mod_TOV, TOV}
\begin{align}
\label{eq:TOV}
\frac{d m}{d r} &= 4 \pi r^{2} \mathcal{E}\, , \nonumber \\
\frac{d P_{r}}{d r} &= -\left(P_{r}+ \mathcal{E}\right)\left(\frac{m}{r^{2}}+4 \pi r P_{r}\right) e^{2 \lambda}+\frac{2}{r} \sigma \, , \nonumber \\
\frac{d \psi}{d r} &= -\frac{1}{P_{r}+ \mathcal{E}} \frac{d P_{r}}{d r}+\frac{2 \sigma}{r\left(P_{r}+ \mathcal{E}\right)} \, , 
\end{align}
where $m(r)$ denotes the enclosed mass corresponding to radius $r$, and $\lambda(r)$ represents the metric function defined as
\begin{equation*}
    e^{-2 \lambda}=1-\frac{2 m}{r}.
\end{equation*}
In order to numerically solve the aforementioned set of coupled ordinary differential equations (ODEs), specific boundary conditions need to be established. Conventionally, the surface of the star is set at $r=R$, where the radial pressure becomes zero ($P_r=0$). As the equilibrium system exhibits spherical symmetry, the Schwarzschild metric is employed to describe the exterior space-time. This choice ensures metric continuity at the surface of the anisotropic neutron star (NS) and imposes a boundary condition on $\psi$. Specifically, the value of $\psi$ at $r=R$ must coincide with the value of $\psi$ in the Schwarzschild metric at $r=R$.
\begin{equation*}
\psi \left(r=R\right)=\frac{1}{2} \ln \left[1-\frac{2  M}{ R}\right] \, .
\end{equation*}
By making a selection of an EOS governing the radial pressure ($P_r$) and adopting an anisotropic model for $\sigma$, it becomes feasible to numerically solve Eq. (\ref{eq:TOV}). This numerical solution involves specifying a central energy density $\mathcal{E}(r=0)=\mathcal{E}_c$, while enforcing the initial condition $m(r=0)=0$.
\begin{figure}
    \centering
    \includegraphics[width=0.5\linewidth]{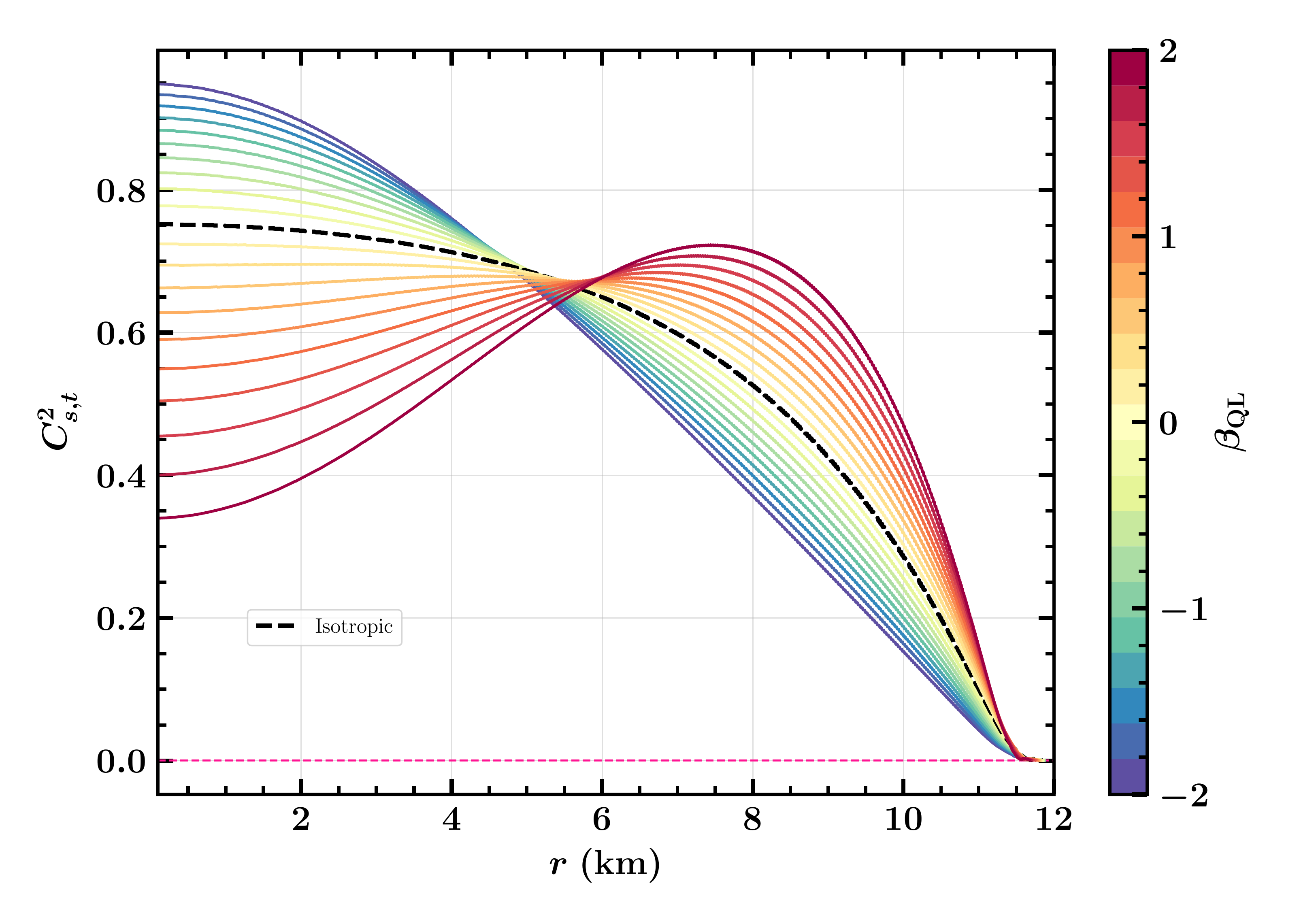}
    \caption{The radial profile of the sound speed ($c_{s,t}^{2}$) for different $\beta_{\rm QL}$ values of the maximum mass NS corresponding to the DD2 EOS.}
    \label{fig:Sound-Speed}
\end{figure}
\subsection{Anisotropy Model}
\label{sec:anisotropi_model}
For anisotropic NS, we use the Quasi-Local (QL) model proposed by Horvat {\it et al.} \cite{QL_Model} describing the quasi-local nature of anisotropy in the following
\begin{equation}
   \sigma = P_{t}-P_{r} = \frac{\beta_{\mathrm{QL}}}{3}P_r \mu = \frac{\beta_{\mathrm{QL}}}{3} P_r (1 - e^{-2\lambda}) \, , 
   \label{eq:QL}
\end{equation}
where the factor $\beta_{\mathrm{QL}}$ depicts the measurement of the degree of anisotropy in the fluid and ($\mu = 1 - e^{-2\lambda} = \frac{2m(r)}{r}$) also known as local compactness is the quasi-local variable. In order to maintain their spherically symmetric configuration, anisotropic NSs must adhere to specific conditions, as outlined in references \cite{Estevez-Delgado2018, Sulaksono}. These conditions include:
\begin{itemize}
\item Absence of anisotropy at the center of the NS i.e. $\sigma = 0$, or equivalently, $P_{r} = P_{t}$ at $r=0$.

\item Positivity of $P_{r}$ and $P_{t}$ throughout the entire star.

\item Positivity of the null energy density $(\mathcal{E})$, dominant energy density $(\mathcal{E}+P_{r}, \mathcal{E}+P_{t})$, and strong energy density $\left(\mathcal{E}+P_{r}+2 P_{t}\right)$ within the star.

\item Non-negativity of the sound speed ($c_s^2$) inside the star, with the $c_s^2$ in the radial and tangential directions satisfying the following constraints: $0 < c_{s,r}^{2}, c_{s, t}^{2} < 1$. It is also essential to ensure that the speed of sound does not exceed the speed of light ($c=1$ in this study).
\end{itemize}
Therefore, the conditions mentioned above are crucial for maintaining the spherical symmetry and physical consistency of anisotropic NSs.

The advantage of using the QL-model with anisotropy is that it ensures the fluid remains isotropic at the center of the star due to the behavior of $(1 - e^{-2\lambda} = 2m/r) \sim r^2$ when $r \rightarrow 0$, while also being applicable only to relativistic configurations where anisotropy may arise at high densities \cite{Herrera2013}. For the 60 EOS-ensembles considered in this study, the QL-model with anisotropy parameter ranging from $-2 < \beta_{\rm QL} < 2$ satisfies all the necessary conditions to maintain spherical symmetry in an anisotropic NS configuration \cite{Estevez-Delgado2018, Sulaksono}. Fig. \ref{fig:Sound-Speed} shows that the speed of sound in the tangential direction ($c_{s,t}^{2}=\frac{\partial P_t}{\partial \mathcal{E}}$) for maximum mass configuration corresponds to DD2 EOS, satisfies the causality condition throughout the star for $-2 < \beta_{\rm QL} < 2$.

\begin{figure}
    \centering
    \includegraphics[width=0.5\linewidth]{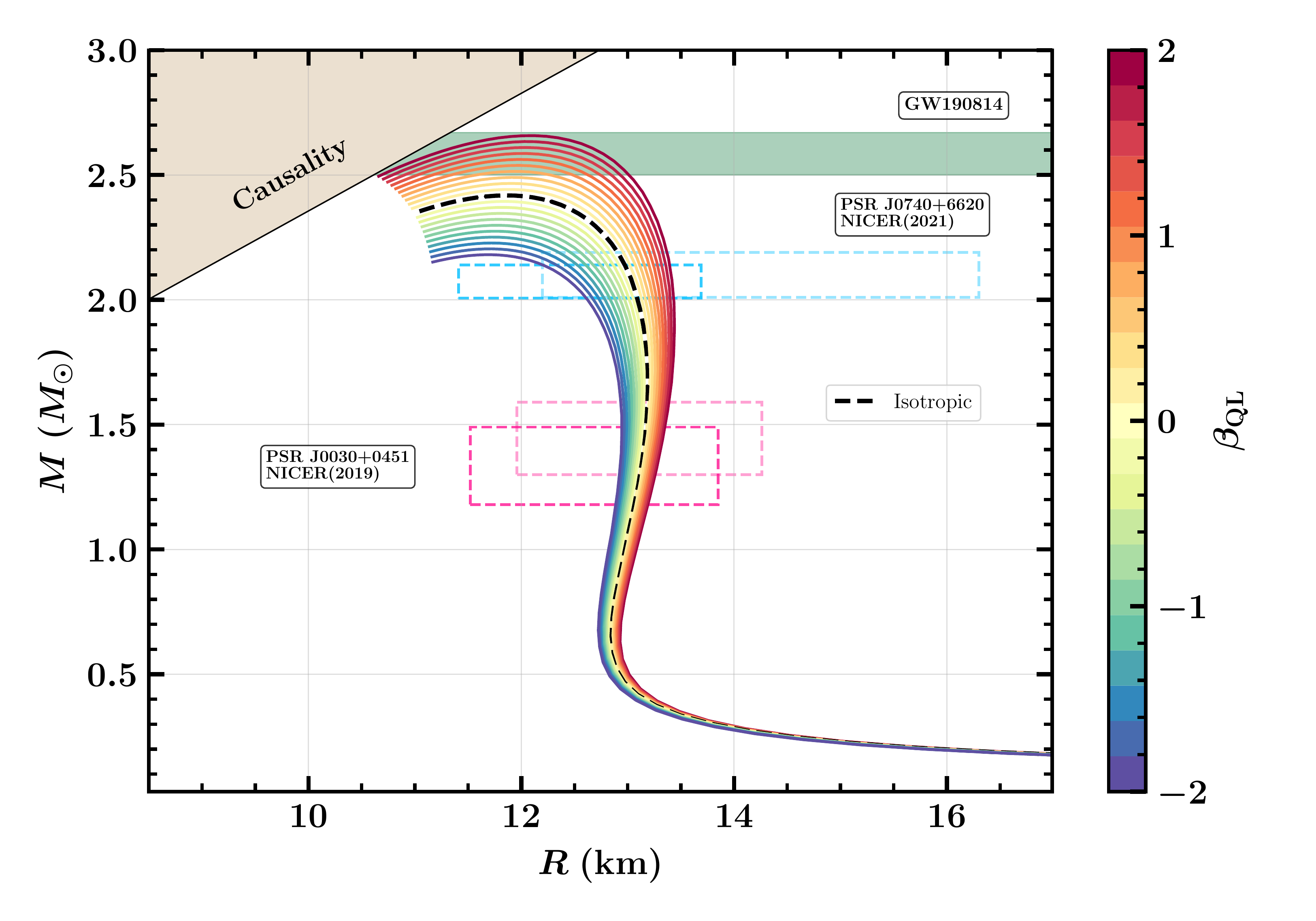}
     \includegraphics[width=0.5\linewidth]{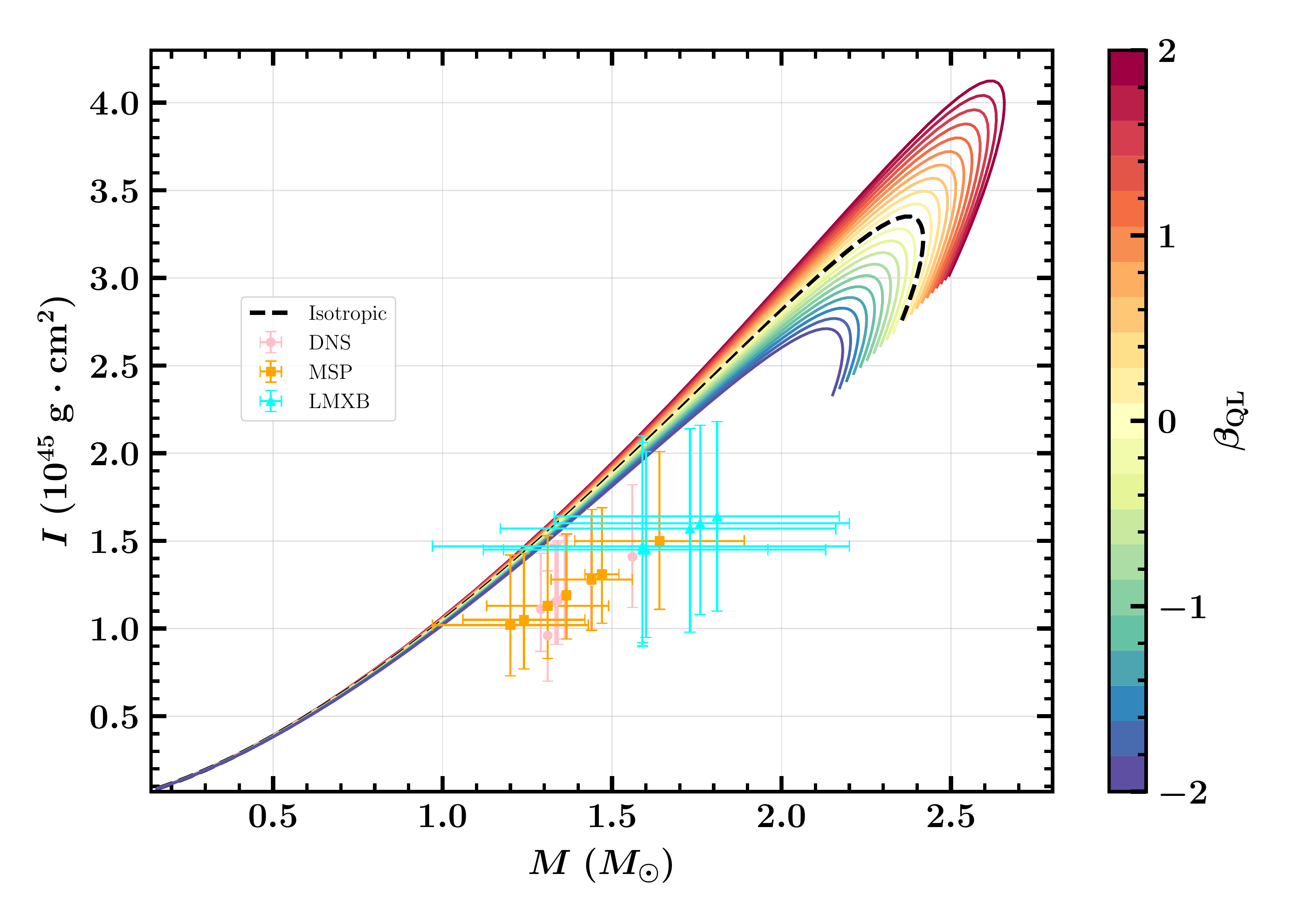}
    \caption{\textit{Left:} Mass-radius profiles for anisotropic NSs with $-2 < \beta_{\rm QL} < 2$ for the DD2 EOS. The black dashed line represents the isotropic case. The limits on mass and radius from the PSR J0030+0451 \cite{Miller_2019, Riley_2019} are shown in the light pink boxes, and the revised NICER data \cite{Miller_2021} is shown in the light blue boxes. The green horizontal bar represents the mass range observed in the GW190814 event \cite{GW190814}. \textit{Right:} The MI as a function of mass for different values of $\beta_{\rm QL}$. The error bars were calculated based on the results of several pulsar analyses as done in Ref. \cite{Bharat_and_Landry}.}
    \label{fig:MR-mom}
\end{figure}

The mass-radius (MR) profile for a given EOS can be obtained by solving the TOV Eqs. (\ref{eq:TOV}) for various central densities, which generate a sequence of mass and radius. Figure \ref{fig:MR-mom} illustrates the MR profiles for the anisotropic star for the DD2 EOS. Adjusting the value of $\beta_{\rm QL}$ influences the maximum mass and the corresponding radius of the NS. The positive value of $\beta_{\rm QL}$ increases the maximum mass and its associated radius, and vice-versa for $\beta_{\rm QL}$. Observational data from different observations, such as X-ray, NICER, and GW (GW170817 and GW190814), to constrain the degree of anisotropy within NS \cite{Miller_2019, Riley_2019, Miller_2021, GW170817, GW190814}. For example, values of $1<\beta_{\mathrm{QL}}<2$ satisfy the mass constraint (2.50–2.67 $M_\odot$) of the GW190814 event, suggesting that one of the merger companions may have been a highly anisotropic NS \cite{Roupas2021}.

\subsection{Slowly Rotating NS and Moment of Inertia}

The MI of a slowly rotating anisotropic NS can be expressed as \cite{Rahmansyah2020, HC_I_LOVE_C}
\begin{equation}
\label{eq:mom}
   I=\frac{8 \pi}{3} \int_{0}^R \frac{r^5 J \bar{\omega}}{r-2 m}\frac{\mathcal{E}+P_r}{\Omega}\left[1+\frac{\sigma}{\mathcal{E}+P_r}\right] dr \, . 
\end{equation}
The MI of an anisotropic NS is shown as a function of its mass in the left panel of Fig. \ref{fig:MR-mom}. As the NS mass increases, the MI also increases until a stable configuration is reached, after which it starts to decrease. Furthermore, both the mass and MI of the NS increase with positive values of $\beta_{\rm QL}$, while the opposite trend is observed for negative values of $\beta_{\rm QL}$. The impact of anisotropy on the MI is more pronounced for high-mass NSs compared to low-mass ones. Kumar and Landry \cite{Bharat_and_Landry} have established constraints on the MI inferred from various sources such as double neutron stars (DNS), millisecond pulsars (MSP), and low-mass X-ray binaries (LMXB). The error bars in the figure represent the possible range of values for these constraints.

\subsection{Tidal Deformability Parameters}
\begin{figure}
    \centering
    \includegraphics[width=0.5\linewidth]{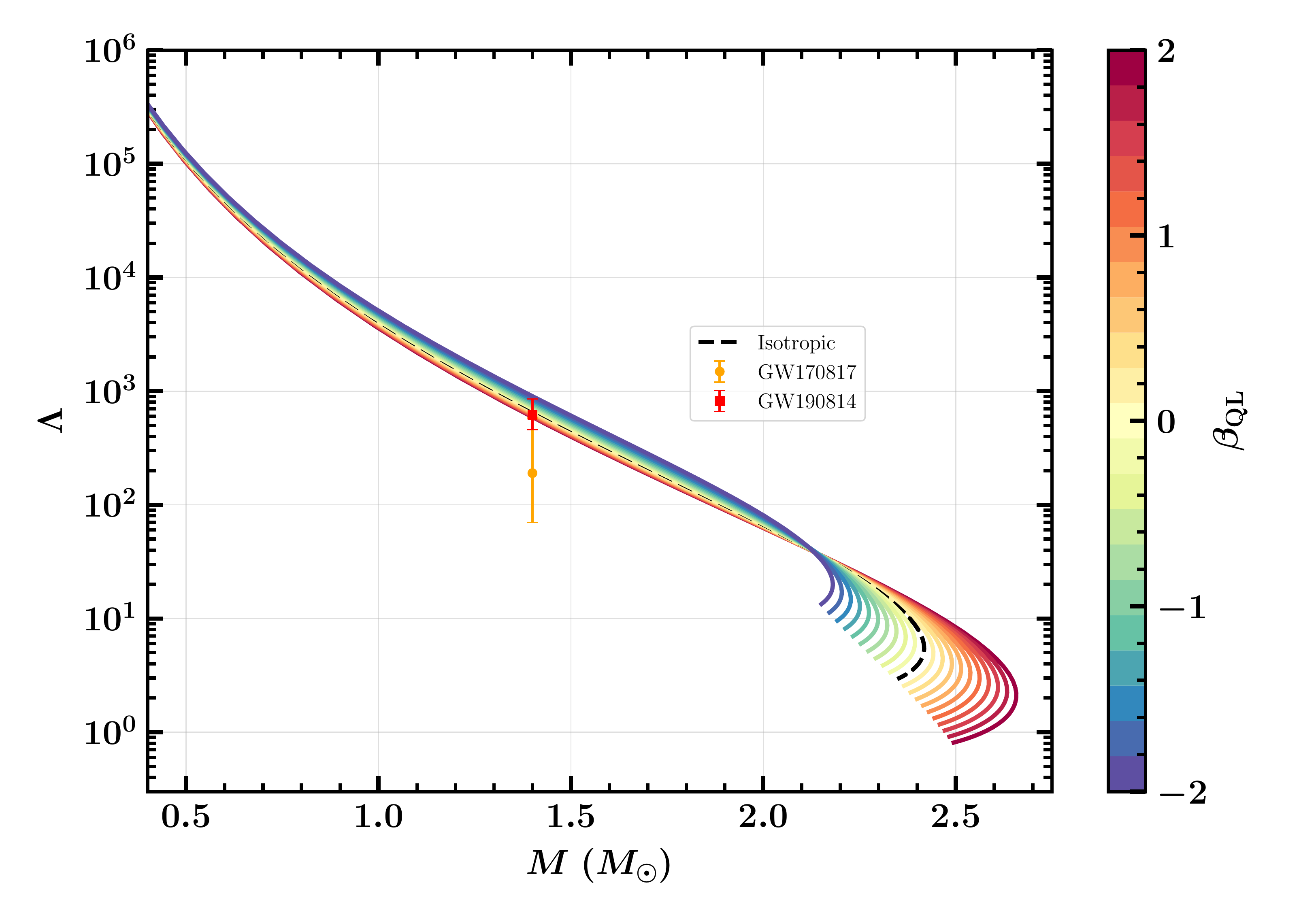}
     \includegraphics[width=0.5\linewidth]{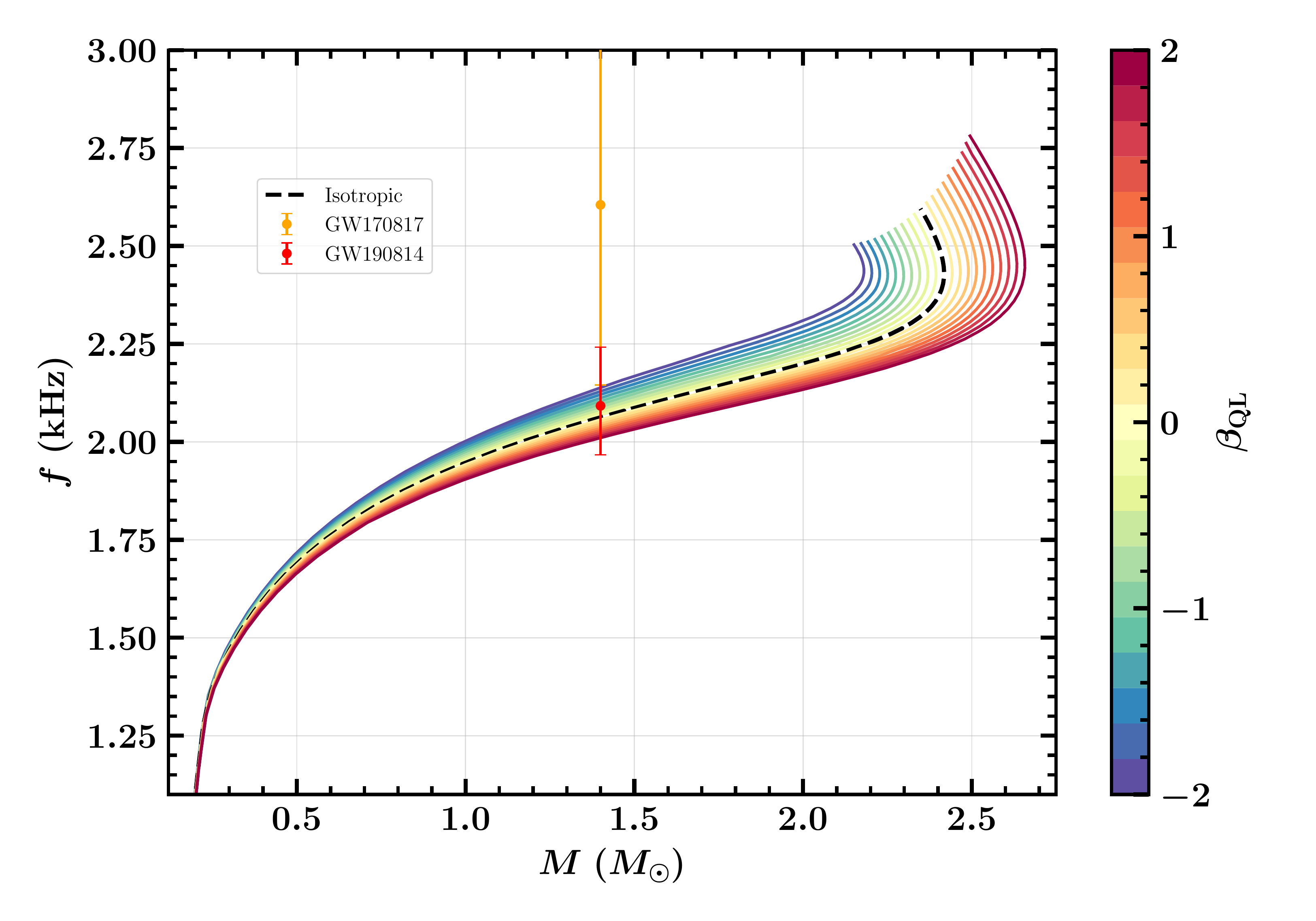}
    \caption{\textit{Left:} The $\Lambda$ as a function of the mass for different values of the $\beta_{\rm QL}$ corresponds to DD2 EOS. The error bars in both panels represent the observational constraints from GW170817 \cite{GW170817} and GW190814 events \cite{GW190814}. \textit{Right:} The $f$-mode frequency as a function mass. The error bars represent the theoretical limits that we obtained in sub-sec. (\ref{constraining_f_mode}).}
    \label{fig:Tidal_fmode}
\end{figure}
When the NS is present in the external field ($\epsilon_{ij}$) created by its companion star, it acquires a quadrupole moment ($Q_{ij}$). The magnitude of the quadrupole moment is linearly proportional to the tidal field and is given by \cite{Hinderer_2008, Hinderer_2009}
\begin{equation}
Q_{i j}=-\alpha \epsilon_{i j},
\end{equation}
where $\alpha$ is the tidal deformability of a star. $\alpha$ can be defined in terms of tidal Love number $k_{2}$ as $\alpha=\frac{2}{3} k_{2} R^{5}$. The dimensionless tidal deformability of the star is defined as $\Lambda = \alpha / M^{5} = \frac{2}{3} k_{2} C^{-5}$. The detailed derivation of $k_2$ for an anisotropic star can be found in Ref. \cite{Bhaskar_Biswas, HC_I_LOVE_C, Sulaksono}.

The dimensionless tidal deformability of anisotropic NSs is shown in Figure \ref{fig:Tidal_fmode}. As the anisotropy parameter $\beta_{\rm QL}$ increases, the magnitude of the Love number $k_2$ and its corresponding tidal deformability $\Lambda$ decrease, while they increase with decreasing $\beta_{\rm QL}$. \SRM{The impact of anisotropy on tidal deformability, as mentioned above, reverses after attaining maximum mass configuration, beyond which the star becomes unstable.} The GW170817 \cite{GW170817} event constrains $\Lambda_{1.4}$ to be $190_{-120}^{+390}$, while GW190814 \cite{GW190814} put a limit of $\Lambda_{1.4}= 616_{-158}^{+273}$ (in the NS-BH scenario). The predicted value of $\Lambda_{1.4}$ satisfies the GW190814 limit for almost all values of $\beta_{\rm QL}$, whereas, for DD2 EOS, the GW170817 limit is met in the range of $0.5<\beta_{\rm QL}<2$. However, $\Lambda$ sharply decreases once the stable configuration is exceeded.
\subsection{Non-Radial Oscillation in Cowling approximation}
The Cowling approximation, initially proposed by Cowling \cite{Cowling_1941} for Newtonian stars and later extended to the case of NSs by McDermott {\it et al.} \cite{mcdermott_1988}. Under this approximation, the metric perturbations are neglected, keeping the space-time metric fixed. We will provide a brief explanation of the derivation of the perturbation equations in the Cowling formalism in the following, while more comprehensive details can be found in \cite{Doneva_2012}. One can obtain the oscillation equations in the Cowling approximation by considering a harmonic time dependence for the perturbation function $W(r, t)=W(r) \mathrm{e}^{i \omega t}$ and $V(r, t)=V(r) \mathrm{e}^{i \omega t}$, where $\omega$ represents the oscillation frequency in the following \cite{Doneva_2012, Curi2022}

\begin{equation}
\begin{aligned}
W^{\prime} = & \  \frac{d \mathcal{E}}{d P_r}\left[\omega^{2} \frac{\mathcal{E}+ P_t}{\mathcal{E}+ P_r}\left(1+\frac{\partial \sigma}{\partial P_r}\right)^{-1} \mathrm{e}^{\lambda-2 \psi} r^{2} V+\psi^{\prime} W\right] -l(l+1) \mathrm{e}^{\lambda} V \\
& -\frac{\sigma}{\mathcal{E}+ P_r}\left[\frac{2}{r}\left(1+\frac{d \mathcal{E}}{d P_r}\right) W +l(l+1) \mathrm{e}^{\lambda} V\right] ,\\
V^{\prime} = & \ 2 V \psi^{\prime}-\left(1+\frac{\partial \sigma}{\partial P_r}\right) \frac{\mathcal{E}+ P_r}{\mathcal{E}+ P_t} \frac{\mathrm{e}^{\lambda}}{r^{2}} W  -\left[\frac{\sigma^{\prime}}{\mathcal{E}+ P_t}+\left(\frac{d \mathcal{E}}{d P_r}+1\right) \frac{\sigma}{\mathcal{E}+ P_t}\left(\psi^{\prime}+\frac{2}{r}\right)\right. \\
& \left.-\frac{2}{r} \frac{\partial \sigma}{\partial P_r}-\left(1+\frac{\partial \sigma}{\partial P_r}\right)^{-1}\left(\frac{\partial^{2} \sigma}{\partial P_r^{2}} P_r^{\prime}+\frac{\partial^{2} \sigma}{\partial P_r \partial \mu} \mu^{\prime}\right)\right] V .
\end{aligned}
\end{equation}

To solve the equations mentioned earlier, it is necessary to consider boundary conditions at the center and surface of the star in the following
\begin{equation}
\omega^{2} \frac{\mathcal{E}+ P_t}{\mathcal{E}+ P_r}\left(1+\frac{\partial \sigma}{\partial P_r}\right)^{-1} \mathrm{e}^{-2 \psi} V  +\left(\psi^{\prime}-\frac{2}{r} \frac{\sigma}{\mathcal{E}+ P_r}\right) \mathrm{e}^{-\lambda} \frac{W}{r^{2}}=0,
\end{equation}
and the boundary condition at the star center $(r=0)$ satisfies
\[\tilde{W}=-l \tilde{V}\]
where, the functions  $\tilde{W}$ and $\tilde{V}$ are defined as $W=\tilde{W} r^{l+1}$ and $V=\tilde{V} r^{l}$.  In this work, we focus on the quadrupolar modes, which correspond to $l=2$. In Fig. \ref{fig:Tidal_fmode}, we show the $f$-mode frequency of a NS as a function of its mass by varying the anisotropic parameter for the DD2 EOS as a representative case. \SRM{For a specific mass NS, the frequency decreases for a positive value of $\beta_{\rm QL}$ while increases for a negative $\beta_{\rm QL}$ till maximum mass is attained, after which the star becomes unstable}. Using the tidal deformability limit from GW170817 and GW190814, one can impose constraints on the canonical $f$-mode frequency for isotropic and anisotropic stars, as discussed in sub-sec. \ref{constraining_f_mode}. We also overlaid the derived theoretical limit on the figure to assess its consistency.
\section{Universal Relations}
The main purpose of UR is to explore the star properties that are difficult to measure through observations. Several URs have already been proposed to estimate the properties of NS, but they are primarily focused on isotropic cases \cite{PhysRevD.101.124006, Breu_2016, PhysRevD.91.044034, Yagi:2013bca, Kent_yagi_2013, staykov2016}. However, very few studies have been dedicated to URs for anisotropic stars, which are more realistic than isotropic ones. Hence, in this study, we aim to explore various types of URs between the moment of inertia, tidal deformability, compactness, and $f$-mode frequency for anisotropic NSs.

Although some known URs for anisotropic NS have been proposed in Refs. \cite{Kent_yagi_2015, Bhaskar_Biswas, HC_I_LOVE_C}, our primary focus will be on the URs between the moment of inertia, $f$-mode frequency, and compactness ($I$-$f$-$C$) as well as the $f$-$\text{Love}$ relation of the anisotropic NS. The NS oscillate with different modes, emitting gravitational waves (GWs). The oscillation frequencies, such as the $f$-mode , $p$-mode, etc., might be detectable in the near future with our terrestrial detectors. However, to interpret these observations effectively, we require prior theoretical knowledge. Therefore, approximate URs for anisotropic NSs hold great significance in astrophysical observations.

Before delving into different URs, it is necessary to normalize/dimensionless certain key parameters of NSs that are required to obtain the URs. In this study, we have used the unit of MI and frequency of the $f$-mode as kg$\cdot$m$^3$ and Hz, respectively. Therefore, these quantities need to be normalized, and their normalized values are given as
\begin{itemize}
    \item Normalised MI ($\eta) = \sqrt{M^3/I}$
    \item Normalised $f$-mode frequency ($\Bar{\omega}) = \omega M$
\end{itemize}
Here, we calculate the URs between $I$-$f$, $C$-$f$, $I$-$C$, and $f$-$\text{Love}$ for anisotropic NSs. For this study, we chose 60 EOSs, as mentioned in the introduction. Regarding anisotropy, we adopt the same QL-model with different degrees of anisotropy, varying from $-2$ to $+2$. Alternatively, one may opt for other models, such as Bower-Liang's, as used in Ref. \cite{HC_I_LOVE_C}. 

\begin{figure}
    \centering 
    \includegraphics[width=0.5\textwidth]{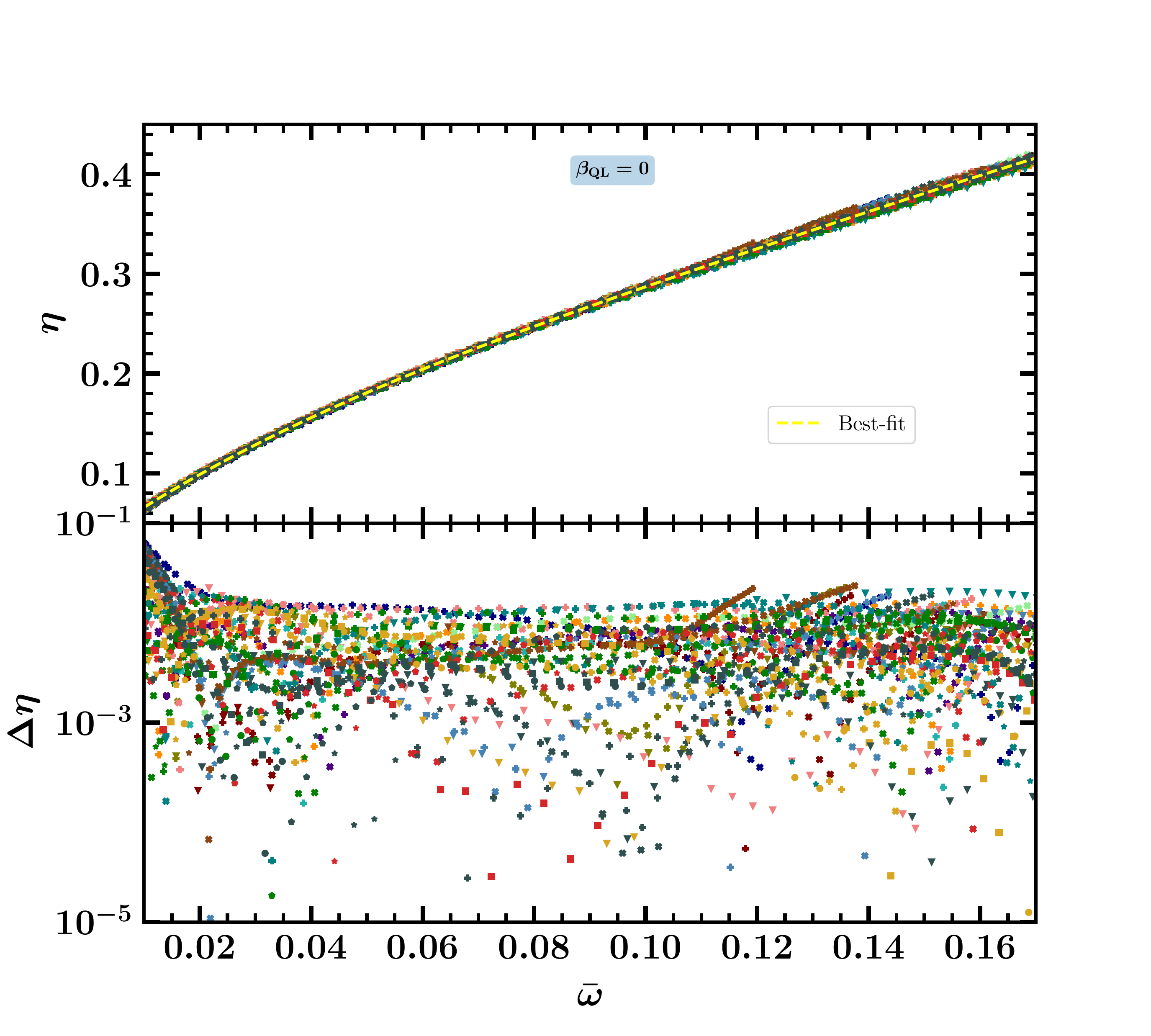}
    \caption{$I$-$f$ relation with anisotropy parameter $\beta_{\rm QL}= 0$ for selected EOSs. The black-dashed line shows the least-squares fit using Eq. (\ref{eq:I-f_fitting}). The lower panel displays the residuals of the fitting obtained using the formula in Eq. (\ref{eq:I-f_residual}).}
    \label{fig:I-f | beta= 0}
\end{figure}
\begin{figure}
    \centering
    \includegraphics[width=0.5\linewidth]{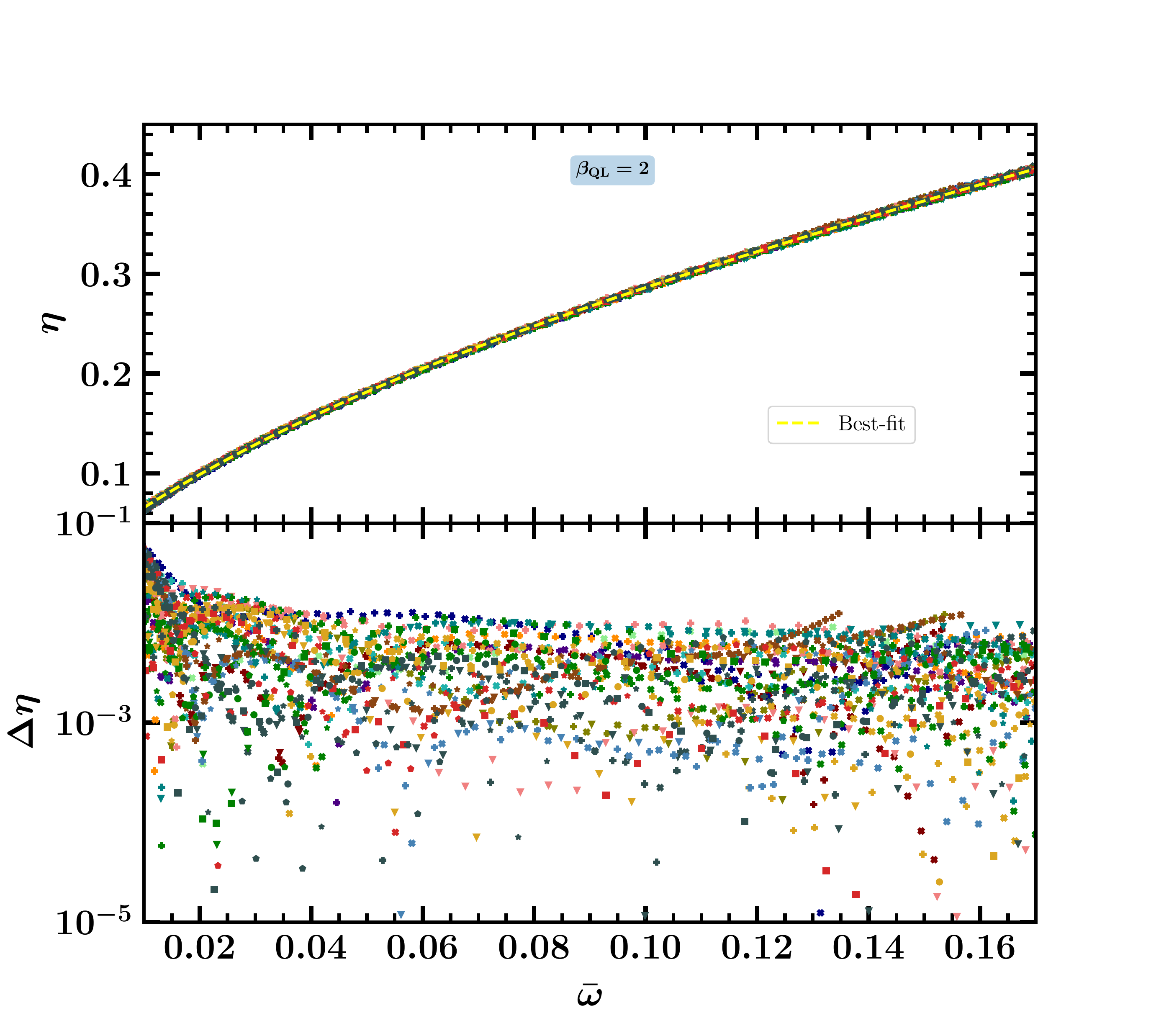}
    \includegraphics[width=0.5\linewidth]{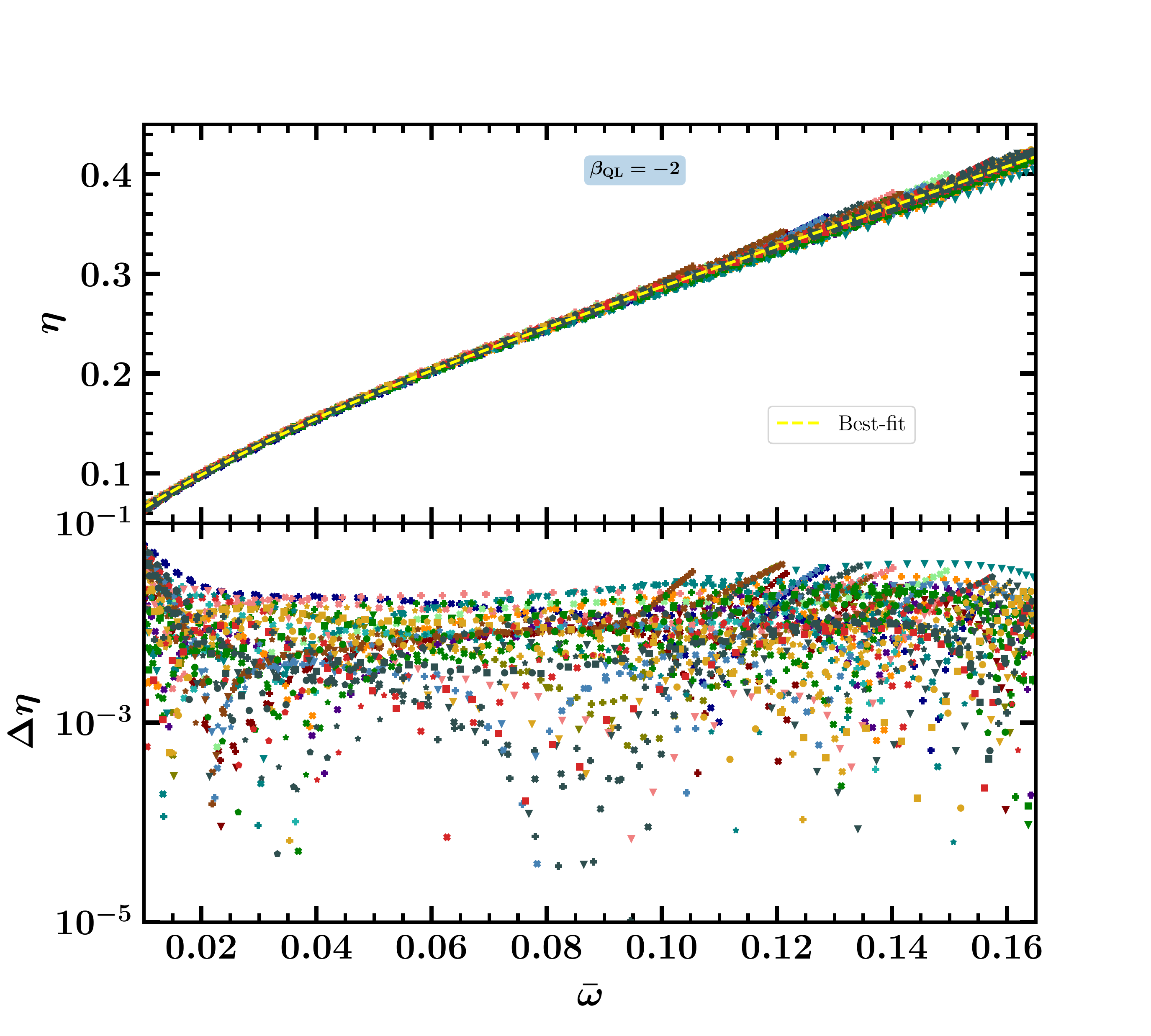}
    \caption{\textit{Left:} Same as Fig. \ref{fig:I-f | beta= 0}, but with $\beta_{\rm QL}= +2$. \textit{Right:} For $\beta_{\rm QL}= -2$}
    \label{fig:I-f | beta= -2}
\end{figure}
\subsection{$I$-$f$ relation}
\begin{figure}
    \centering
    \includegraphics[width=0.5\linewidth]{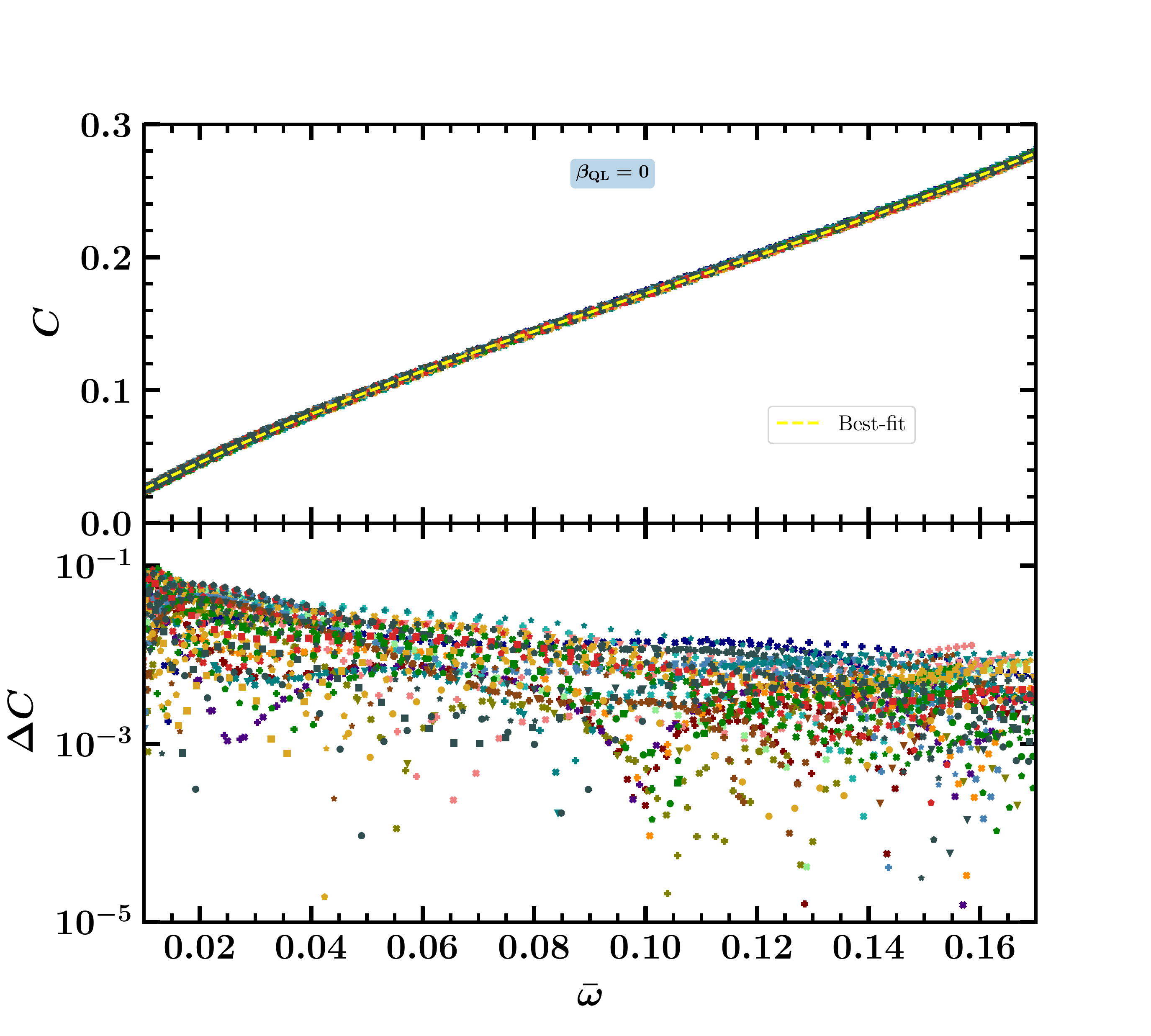}
    \caption{$C$-$f$ relation with anisotropy parameter $\beta_{\rm QL}= 0$ for assumed EOSs. The black-dashed line is fitted with Eq. (\ref{eq:C-f_fitting}). The lower panel shows the residuals for the fitting are calculated.}
    \label{fig:C-f | beta= 0}
\end{figure}
\begin{figure}
    \centering
    \includegraphics[width=0.5\linewidth]{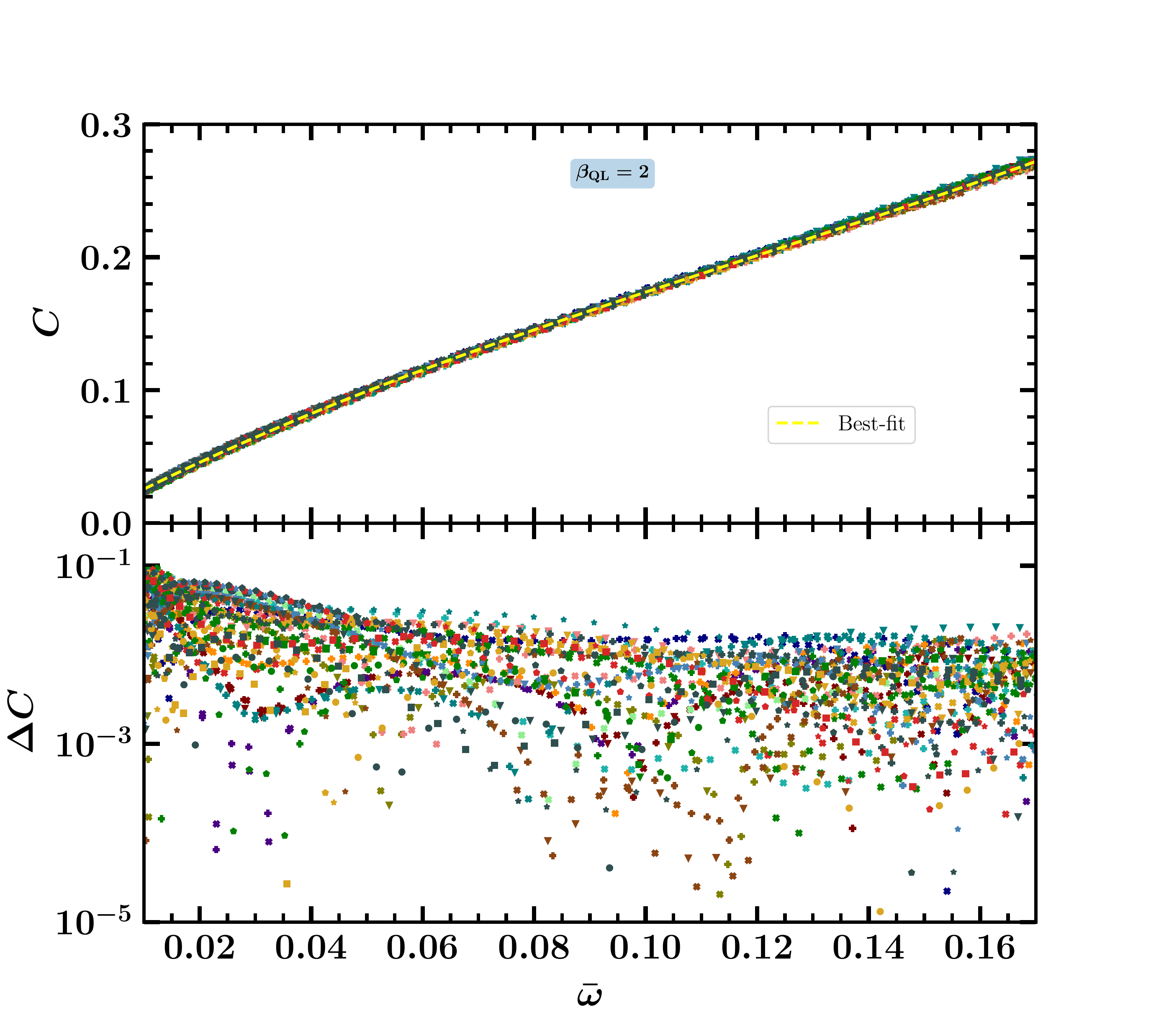}
    \includegraphics[width=0.5\linewidth]{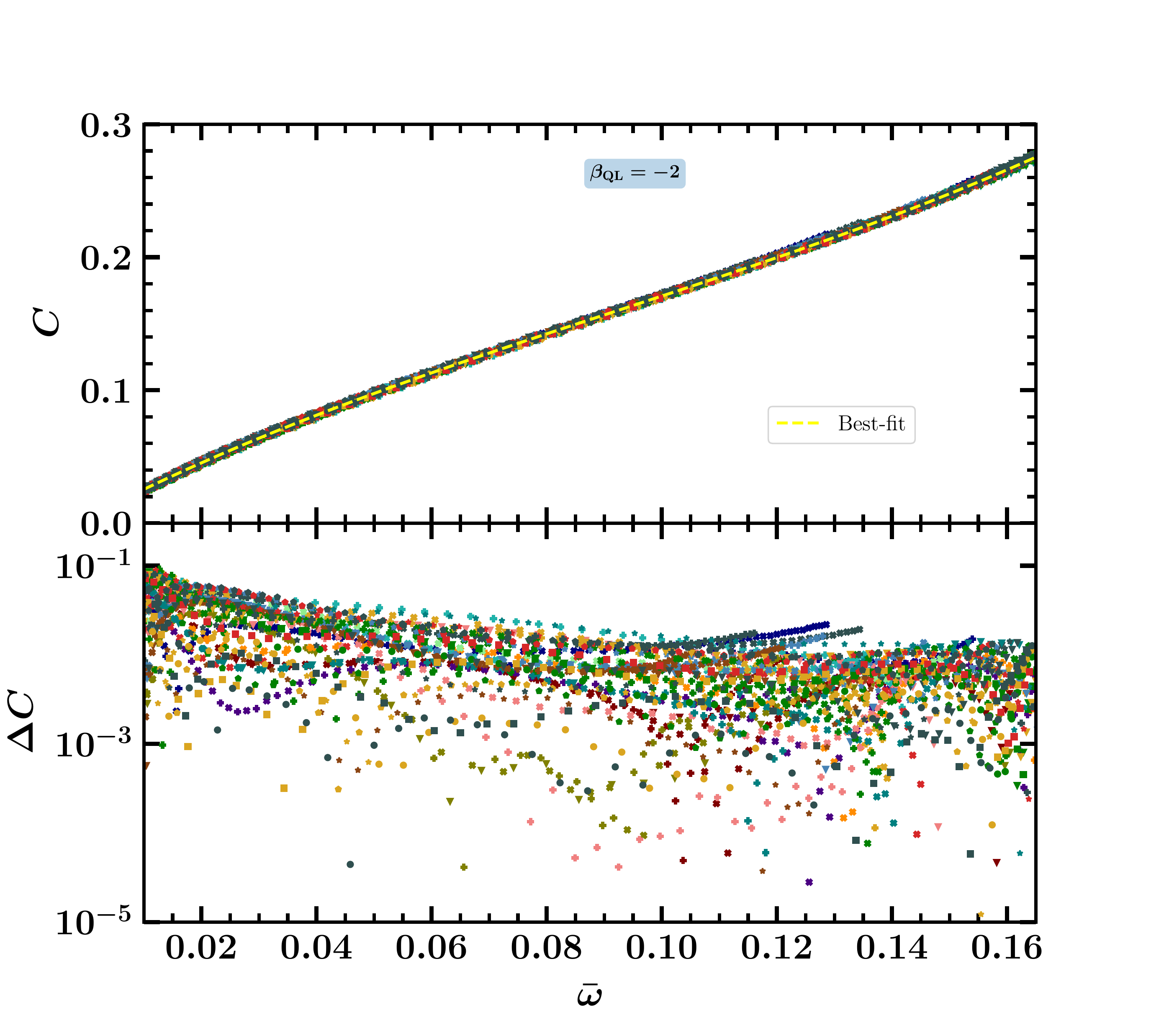}
    \caption{\textit{Left:} Same as Fig. \ref{fig:C-f | beta= 0}, but with $\beta_{\rm QL}= +2$. \textit{Right:} For $\beta_{\rm QL}= -2$}
    \label{fig:C-f | beta= -2}
\end{figure}

The $I$-$f$ UR for isotropic NS with a few EOSs was first calculated by Lau {\it et al.} \cite{Lau_2010}. Breu and Rezzolla \cite{Breu_2016} studied the universal behavior of dimensionless MI, which is defined as $\Bar{I}=I/M^3$, and is more accurate than the dimensionless MI defined earlier, $\Bar{I}=I/MR^2$. Lau and Leung replaced $\Bar{I}=I/M^3$ with $\eta= \sqrt{M^3/I}$, called as normalized MI/effective compactness, due to its proportionality with compactness for stars. Therefore, in this study, we use $\eta= \sqrt{M^3/I}$ rather than $\Bar{I}=I/MR^2$. The relation between $I$-$f$ for anisotropic NSs is performed using the least-squares fit with the approximate formula
\begin{equation}
\label{eq:I-f_fitting}
    \eta = \sum_{n=0}^{n=4} a_n (\Bar{\omega})^n \, .
\end{equation}
The normalized MI ($\eta$) is plotted as the function of normalized $f$-mode frequency ($\Bar{\omega}$) in Figs. \ref{fig:I-f | beta= 0}-\ref{fig:I-f | beta= -2} with $\beta_{\rm QL}= -2, 0,+2$ for anisotropic NS. The residuals are computed with the formula,
\begin{equation}
\label{eq:I-f_residual}
    \Delta \eta = \frac{\eta - \eta_{\rm fit}}{\eta_{\rm fit}} \, .
\end{equation}
We enumerated the coefficients ($a_n$) with their corresponding reduced chi-squared ($\chi_r^2$) errors in Table \ref{tab:I-f-C_coef}. An increase in anisotropy results in a decrease in the value of $\chi_r^2$ error, indicating stronger EOS insensitive relation, and vice versa.

\subsection{$C$-$f$ relation}
Andersson and Kokkotas \cite{Kokkotas1999} first established the correlation between $C$ and $f$-mode frequency. Here, we calculate the $C$-$f$ relations for anisotropic NSs, using the approximate formula obtained through least-squares fitting
\begin{equation}
\label{eq:C-f_fitting}
    C = \sum_{n=0}^{n=4} b_n (\Bar{\omega})^n \, .
\end{equation}
Compactness is plotted as the function of normalized $f$-mode frequency ($\Bar{\omega}$) in Figs. \ref{fig:C-f | beta= 0}-\ref{fig:C-f | beta= -2} with $\beta_{\rm QL}= -2, 0, +2$ for anisotropic NS. 
The coefficients ($b_n$) with $\chi_r^2-$error are enumerated in Table \ref{tab:I-f-C_coef}. The magnitude of $b_n$ increases with increasing $\beta_{\rm QL}$, implying that the fitting is more robust for the isotropic case. Additionally, $\chi_r^2$ also increases with the inclusion of anisotropy, and it is the minimum for the isotropic case. Therefore, the inclusion of anisotropy (whether positive or negative) weakens the EOS insensitive $C$-$I$ UR.
\begin{table}
\caption{The fitting coefficients are listed for $I$-$f$, $C$-$f$, and $C$-$I$ relations with $\beta_{\mathrm{QL}}=$ -2.0, -1.0, 0.0, +1.0,  +2.0. The reduced chi-squared ($\chi_r^2$) is also given for all cases.}
\centering
\setlength{\tabcolsep}{1.7pt}
\renewcommand{\arraystretch}{1.5}
\scalebox{0.9}{
    \begin{tabular}{cccccc}
        \hline \hline
        \multicolumn{6}{c}{$I$-$f$}  \\
        \hline
        $\beta_{\mathrm{QL}}=$ & -2.0 & -1.0 & 0.0 & +1.0 & +2.0   \\
        \hline
        $a_0\left(10^{-2}\right)=$ & 2.825 & 2.848 & 2.918 & 2.886 & 2.903 \\
        $a_1=$ & 3.945 & 3.919 & 3.865 & 3.874 & 3.856   \\
        $a_2\left(10^{1}\right)=$ & -2.436 & -2.303 & -2.139 & -2.092 & -2.011  \\
        $a_3\left(10^{2}\right)=$ & 1.364 & 1.218 & 1.059 & 0.982 & 0.889 \\
        $a_4\left(10^{2}\right)=$ & -2.852 & -2.457 & -2.064 & -1.852 & -1.613 \\
        $\chi_r^2\left(10^{-6}\right)=$ & 13.511 & 8.514 & 5.338 & 3.541 & 2.517  \\
        \hline \hline
    \end{tabular} \hspace{2mm}
    \begin{tabular}{cccccc}
        \hline \hline
        \multicolumn{6}{c}{$C$-$f$}  \\
        \hline
         $\beta_{\mathrm{QL}}=$ & -2.0 & -1.0 & 0.0 & +1.0 & +2.0   \\
        \hline
         $b_0\left(10^{-3}\right)=$ & 4.007 & 4.093 & 4.084 & 4.197 & 4.249 \\
         $b_1=$ & 2.232 & 2.220 & 2.223 & 2.212 & 2.209  \\
         $b_2=$ & -8.752 & -8.143 & -7.963 & -7.531 & -7.269 \\
         $b_3\left(10^{1}\right)=$ & 3.066 & 2.656 & 2.629 & 2.371 & 2.211 \\
        $b_4=$ & 4.836 & 6.740 & -3.179 & 5.173 & -7.423 \\
         $\chi_r^2\left(10^{-6}\right)=$ & 2.228 & 1.702 & 1.846 & 2.282 & 2.871  \\
        \hline \hline
    \end{tabular}} \vspace{3mm}  \\
\scalebox{0.9}{
    \begin{tabular}{cccccc}
        \hline \hline
        \multicolumn{6}{c}{$C$-$I$}  \\
        \hline
          $\beta_{\mathrm{QL}}=$ & -2.0 & -1.0 & 0.0 & +1.0 & +2.0  \\
        \hline
        $c_0\left(10^{-3}\right)=$ & -8.651 & -8.244 & -8.941 & -7.860 & -7.637 \\
        $c_1\left(10^{-1}\right)=$ & 4.473 & 4.348 & 4.477 & 4.236 & 4.171 \\
        $c_2=$ & 1.360 & 1.465 & 1.389 & 1.557 & 1.612 \\
        $c_3=$ & -3.986 & -4.288 & -4.080 & -4.516 & -4.668 \\
        $c_4=$ & 4.892 & 5.272 & 5.147 & 5.610 & 5.827 \\
       $\chi_r^2\left(10^{-6}\right)=$ & 6.338 & 6.382 & 6.436 & 6.313 & 6.210 \\
        \hline \hline
    \end{tabular}}
    \label{tab:I-f-C_coef}
\end{table}

\begin{figure}
    \centering
    \includegraphics[width=0.47\linewidth]{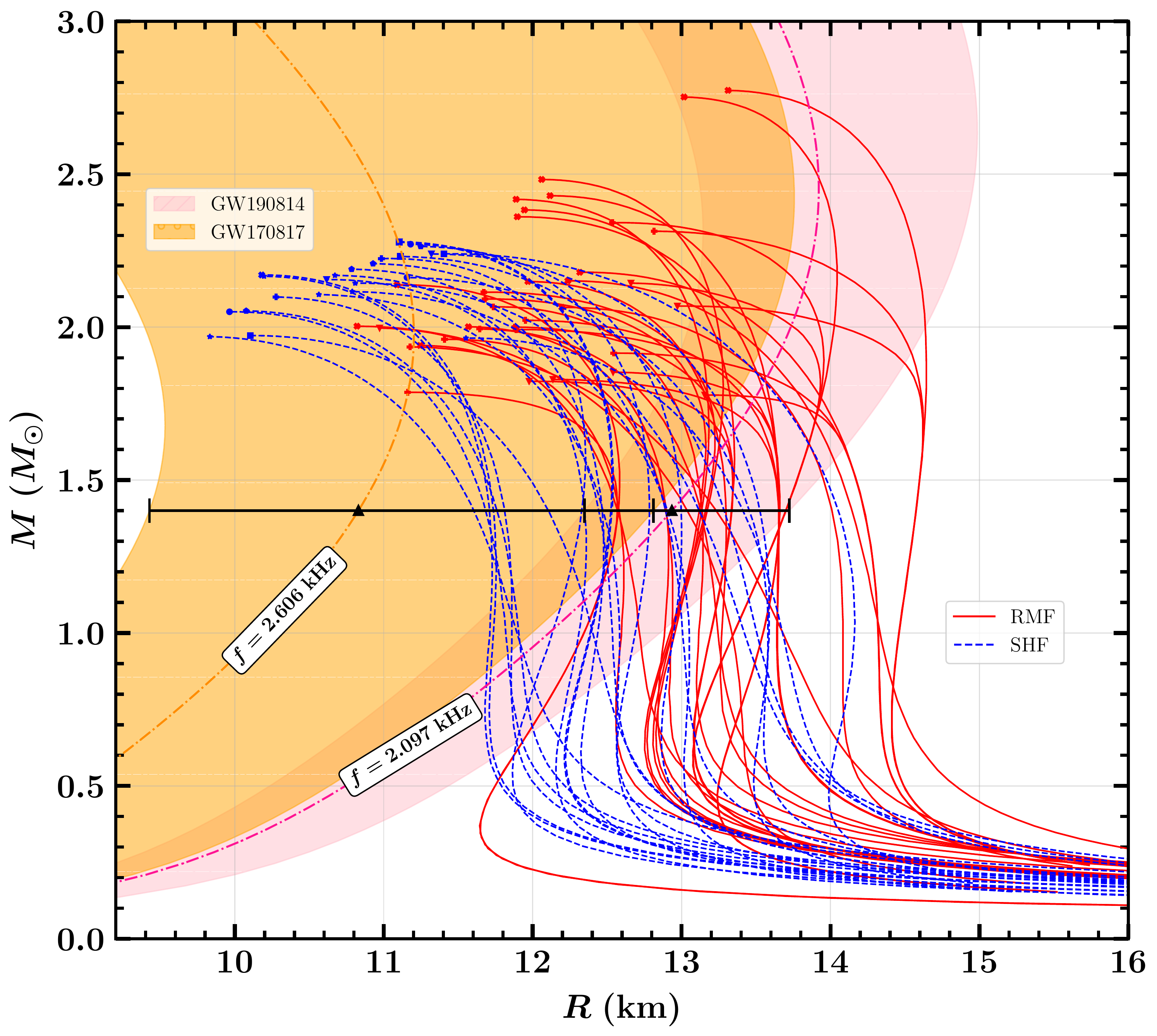}
    \includegraphics[width=0.5\linewidth, height=0.42\linewidth]{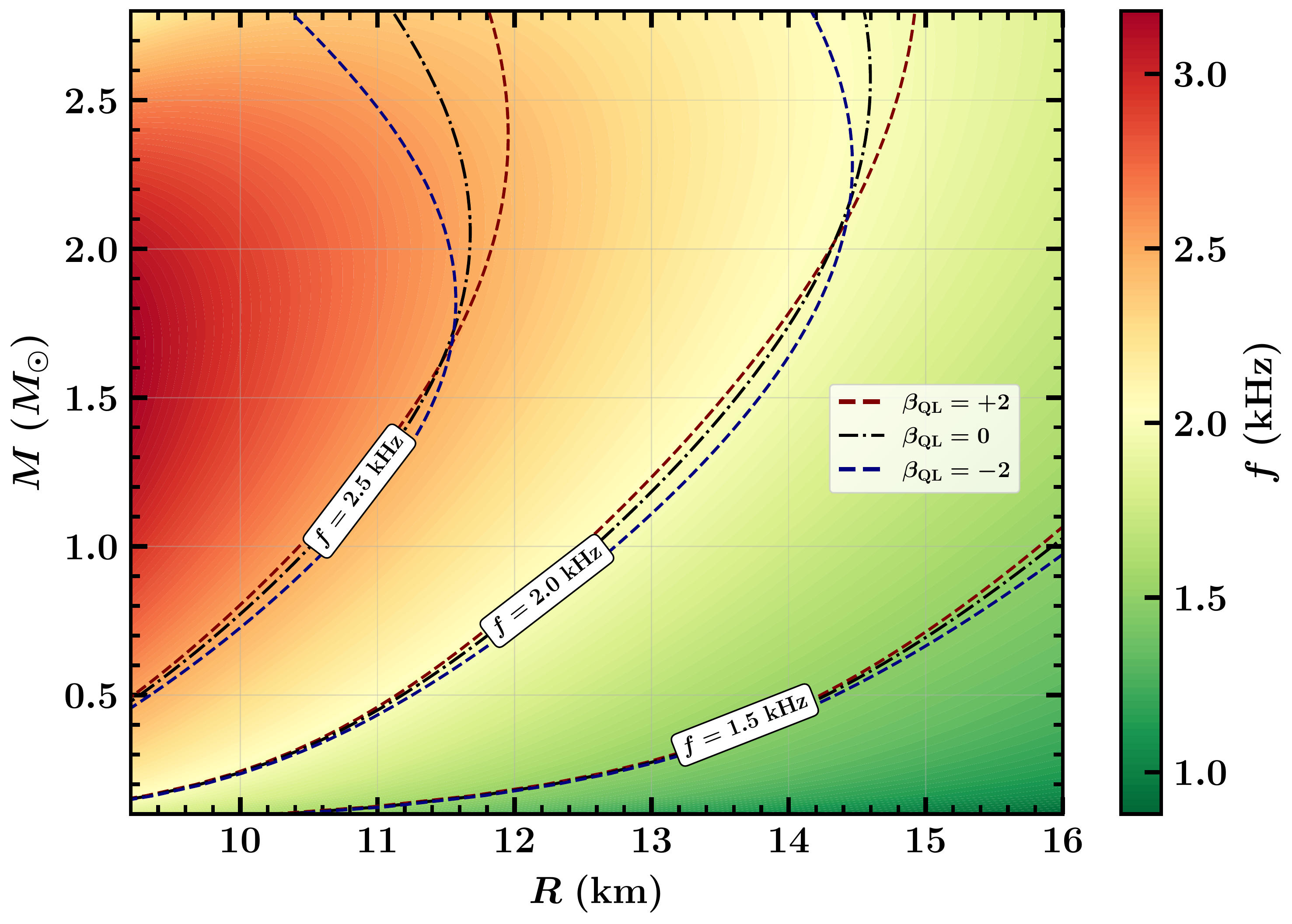}
    \caption{\textit{Left:} Mass-radius profiles of isotropic NSs for EOS ensembles that we have considered in this study. The orange and pink colored MR bands correspond to limits on NS's mass and radius imposed by isotropic $C$-$f$ UR for the canonical $f$-mode frequency ($f_{1.4}$) that was obtained through $f-$Love UR with the help of tidal deformability constraints ($\Lambda_{1.4}$) of GW170817 \cite{GW170817} and GW190814 \cite{GW190814} events, respectively. The horizontal error bars illustrate the radius limits for a canonical mass NS imposed by the frequency bands for respective events. \textit{Right:} The frequency distribution contour plot across the $M$-$R$ parameter space. This distribution is imposed by the isotropic $C$-$f$ UR. $M$-$R$ lines for isotropic cases corresponding to a set of frequencies are shown in black dashed-dot lines. The red and blue dashed lines represent the $M$-$R$ curves for anisotropy parameter $\beta_{\rm QL} =+2$  and $-2$, respectively.}
    \label{fig:MR_Bandandcontour}
\end{figure}

\SRM{One of the primary applications of $C$-$f$ UR involves determining $M$ and $R$ based on the analysis of observed mode data, as articulated by Andersson and Kokkotas \cite{Andersson_1998}. For a unique choice of $f$-mode frequency the $C$-$f$ UR can be exploited to construct a $M$-$R$ relation. This constrained relationship, accounting for uncertainties represented by standard deviations in UR, yields $M$-$R$ bands, illustrated in the left panel of Fig. \ref{fig:MR_Bandandcontour}. In this representation, the orange $M$-$R$ band delineates a region where neutron stars are anticipated to exhibit a frequency of $f= 2.606 ^{+0.457} _ {-0.484}$ kHz, with the solid dashed line denoting $f= 2.606$ kHz. Similarly, the pink band corresponds to a region where neutron stars are expected to possess a frequency of $f= 2.097 ^{+0.124} _ {-0.149}$ kHz, and the solid dashed line represents $f= 2.097$ kHz. It is noteworthy that the frequency constraints employed for plotting $M$-$R$ bands align with the canonical $f$-mode frequency constraints for isotropic neutron stars determined in this study for the GW170817 and GW190814 events. The horizontal error bars in the left panel of Fig. \ref{fig:MR_Bandandcontour} indicate radius limits imposed by the $M$-$R$ bands of the respective events, considering a canonical mass neutron star.}

\SRM{The right panel of Fig. \ref{fig:MR_Bandandcontour} portrays the distribution of $f$-mode frequencies across the $M$-$R$ parameter space for isotropic neutron stars based on $C$-$f$ UR. The black dashed line represents a specific set of mass and radius values for isotropic stars, anticipated to exhibit the mentioned frequency according to $C$-$f$ UR. The figure also depicts variations in these $M$-$R$ lines resulting from the inclusion of anisotropy. NSs with frequencies $f<1.5$ kHz lie in the low compactness region and suffer minimal changes in mass and radius due to the inclusion of anisotropy. Through observing $M$-$R$ lines as depicted in Fig. \ref{fig:MR_Bandandcontour}, we can conclude that for a constant mass NS having a fixed frequency with $f \geq 1.5$ kHz, the radius would tend to decrease with the presence of positive anisotropy and increase for negative anisotropy, altering the compactness of the star in order to maintain its natural frequency till a certain critical point/set of mass and radius is reached. After this, the effect of anisotropy on $M$-$R$ lines reverses. \SRM{}\HCD{This kind of behavior in which my effects of anisotropy on the NS parameter reverses occurs due to the presence of an unstable core which suggests us that the critical mass-radius point in the $M$-$R$ curves is the maximum stability point, beyond which the NSs are unstable in nature.}}
\begin{figure}
    \centering
    \includegraphics[width=0.5\linewidth]{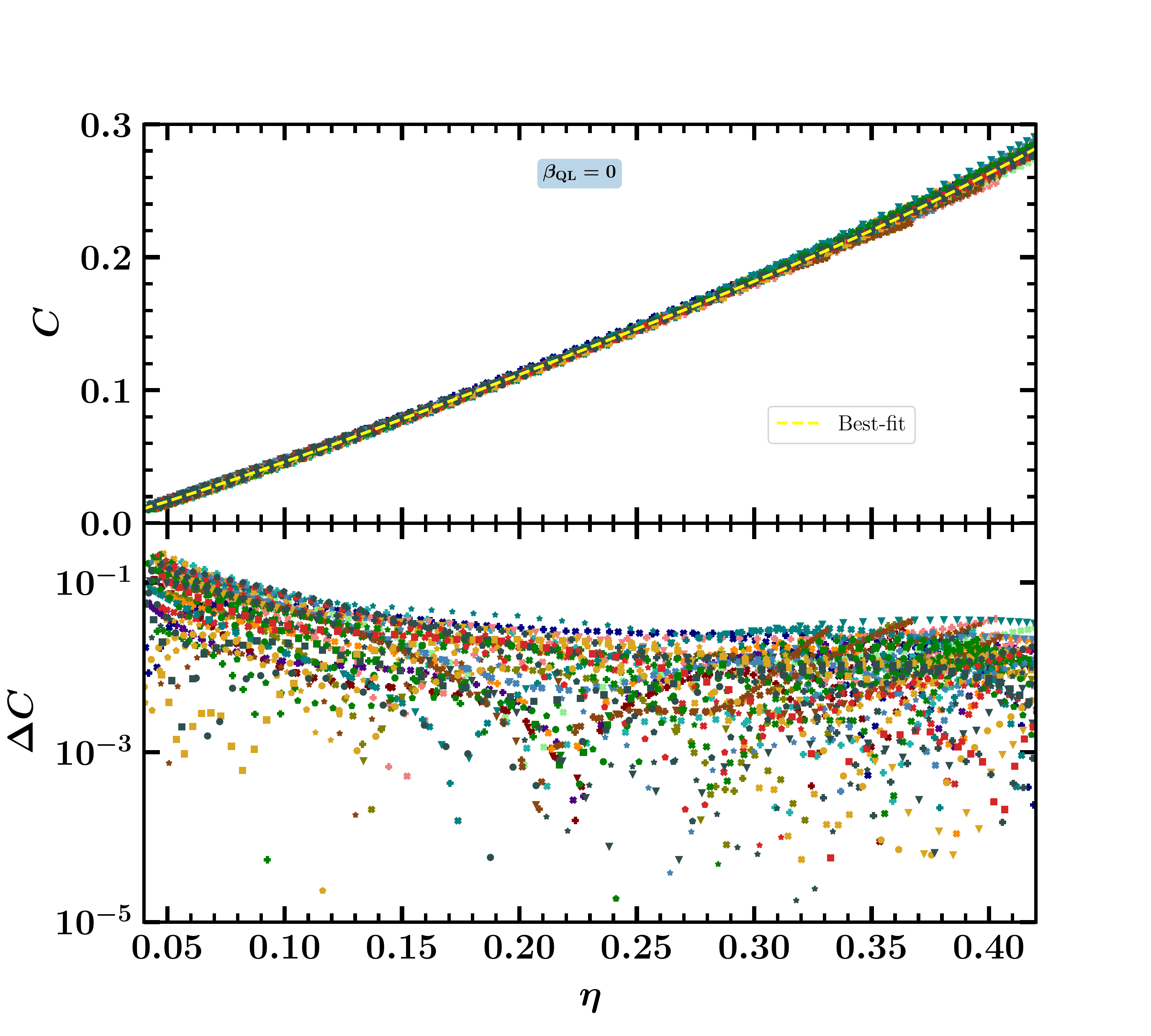}
    \caption{$C$-$I$ relation with anisotropy parameter $\beta_{\rm QL}= 0$ for assumed EOSs. The black-dashed line is fitted with Eq. (\ref{eq:C-I_fitting}). The lower panel shows the residuals for the fitting are calculated.}
    \label{fig:C-I | beta= 0}
\end{figure}
\begin{figure}
    \centering
    \includegraphics[width=0.5\linewidth]{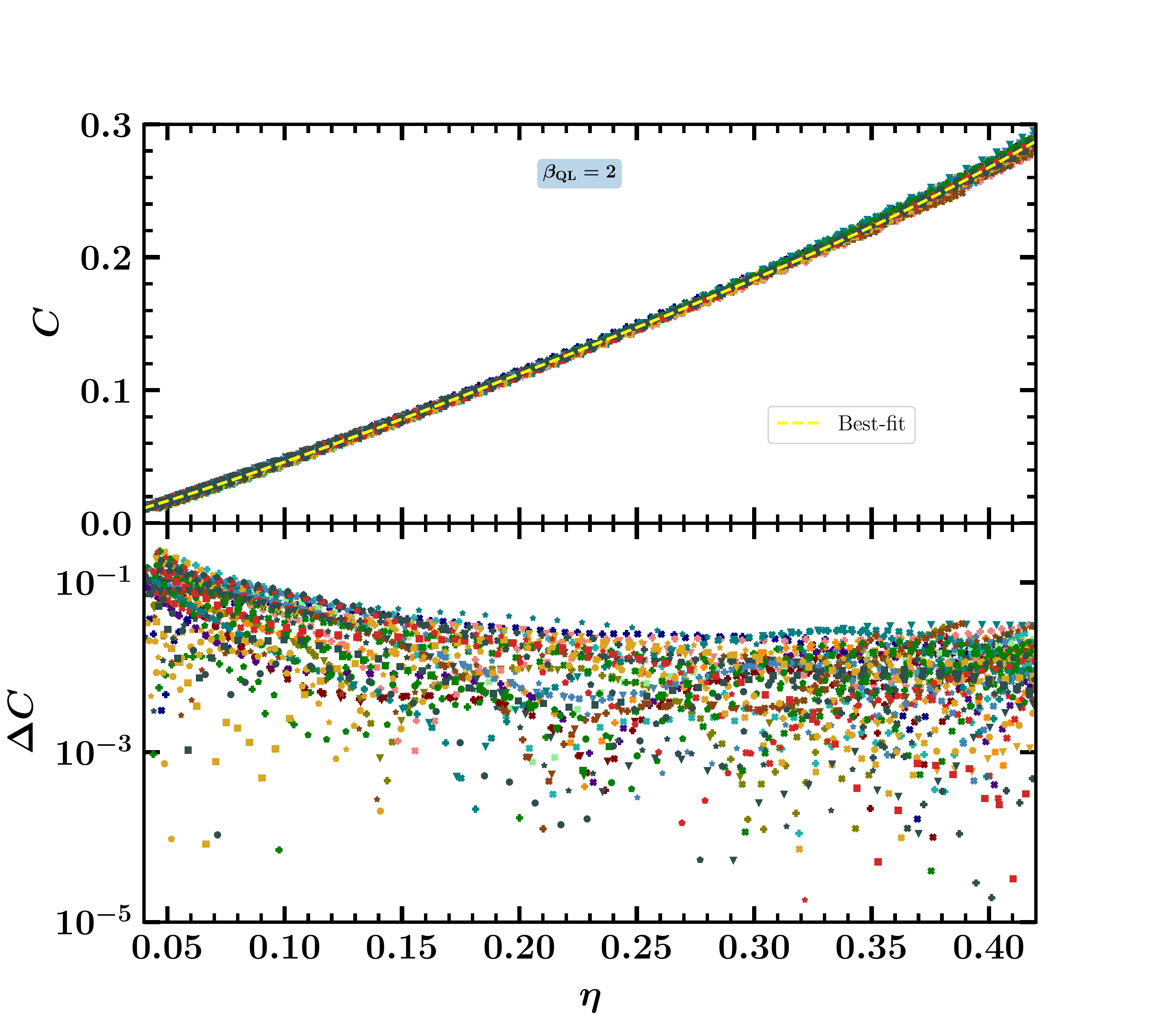}
    \includegraphics[width=0.5\linewidth]{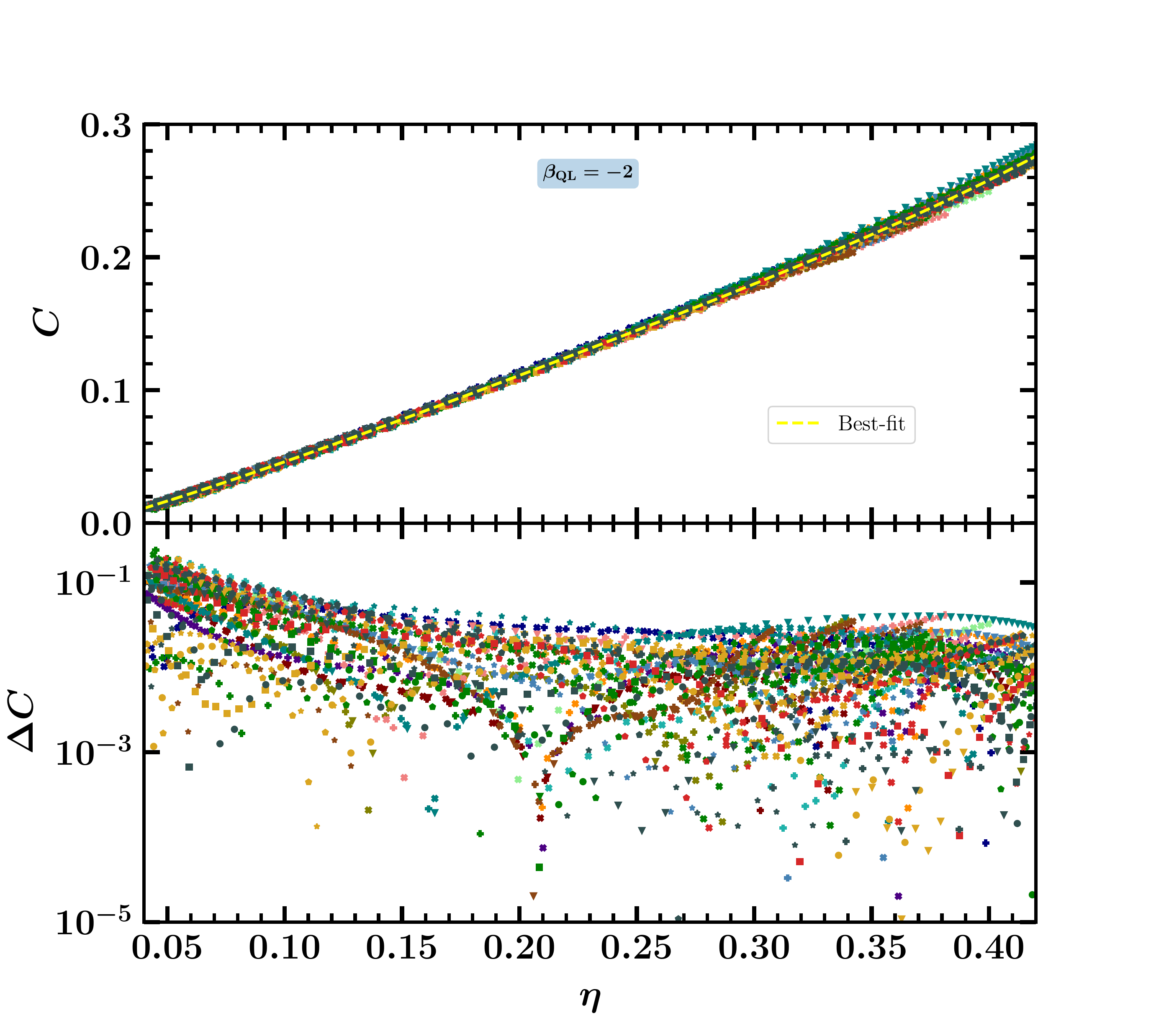}
    \caption{\textit{Left:} Same as Fig. \ref{fig:C-I | beta= 0}, but with $\beta_{\rm QL}= +2$. \textit{Right:} Same as Fig. \ref{fig:C-I | beta= 0}, but with $\beta_{\rm QL}= -2$}
    \label{fig:C-I | beta= -2}
\end{figure}

\subsection{$C$-$I$ relation}
The relationship between the dimensionless MI ($\Bar{I}= I/M^3$) and compactness has been established as a lower-order polynomial fit by Ravenhall and Pethick \cite{Ravenhall_1994}. Since then, this relation has been studied and modified by various authors, including for the double pulsar system with higher-order polynomial fitting \cite{Lattimer_2005}, scalar-tensor theory and $R^2$ gravity \cite{staykov2016, Popchev2019}, rotating stars \cite{Breu_2016}, and strange stars \cite{refId0}. In this work, we investigate the $C$-$I$ relations for anisotropic NS using the normalized moment of inertia ($\eta= \sqrt{M^3/I}$) instead of the dimensionless one. We use the approximate formula to perform a least-squares fit
\begin{equation}
\label{eq:C-I_fitting}
    C = \sum_{n=0}^{n=4} c_n (\eta)^n .
\end{equation}
We display the relationship between $C$ and $\eta$ for anisotropic NSs with $\beta_{\rm QL}= -2, 0,+2$, respectively in Figs. \ref{fig:C-I | beta= 0}-\ref{fig:C-I | beta= -2}. We observe that the inclusion of anisotropy has little effect on the $\chi_r^2$ error, indicating that the $C$-$I$ UR is conserved even when anisotropy is present, especially for NSs with low compactness.
\begin{figure}
    \centering
    \includegraphics[width=0.5\linewidth]{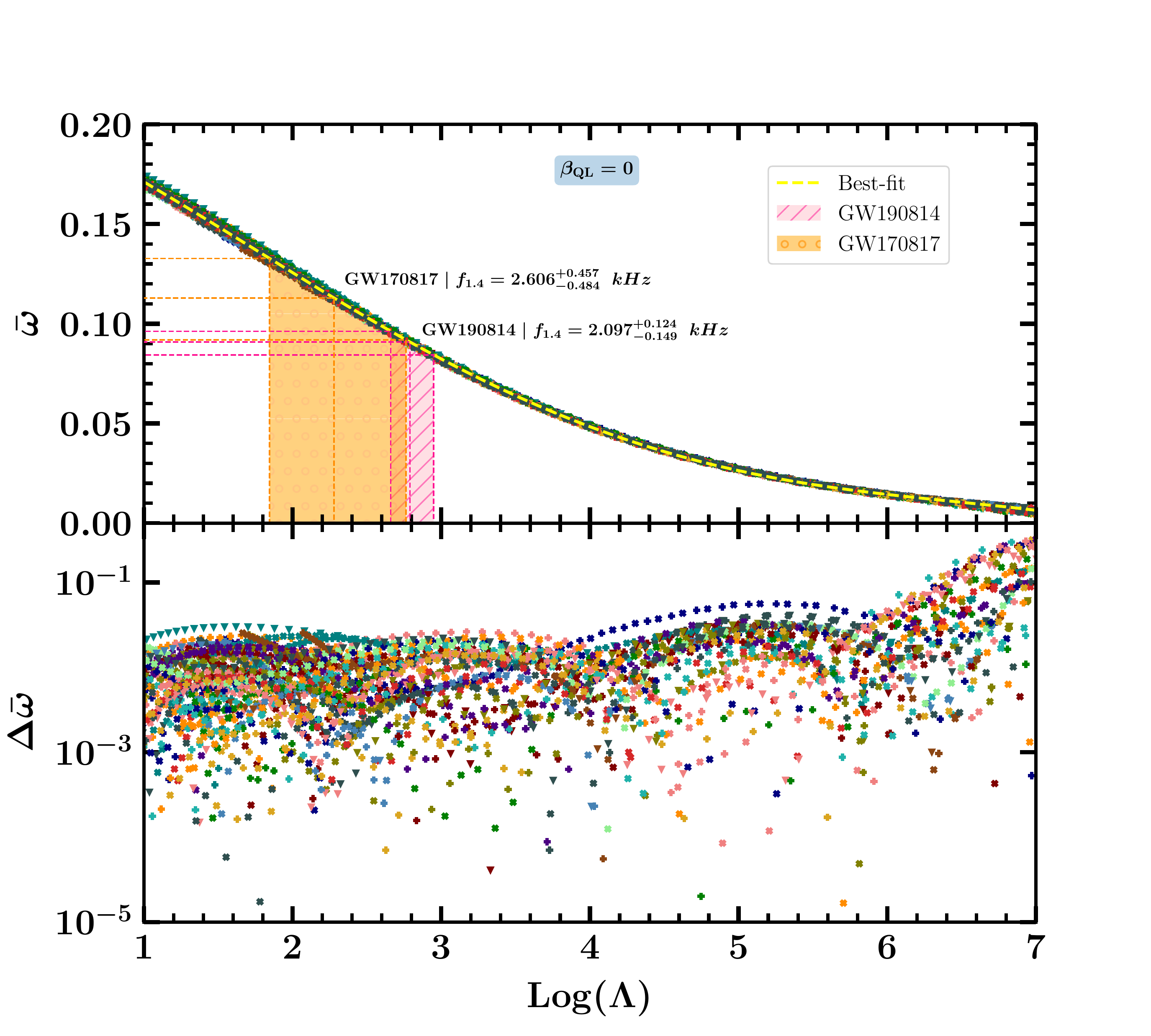}
    \caption{The $f$-Love relation for anisotropic NSs with $\beta_{\rm QL}= 0$ for various assumed EOSs. The black-dashed line represents the best fit using Eq. (\ref{eq:f-Love_fitting}). The light-pink-shaded region and the orange-shaded region represent the range of canonical tidal deformability data obtained from the GW190814 \cite{GW190814} and GW170817 \cite{GW170817} papers, respectively.}
    \label{fig:f-Love | beta= 0}
\end{figure}
\begin{figure}
    \centering
    \includegraphics[width=0.5\linewidth]{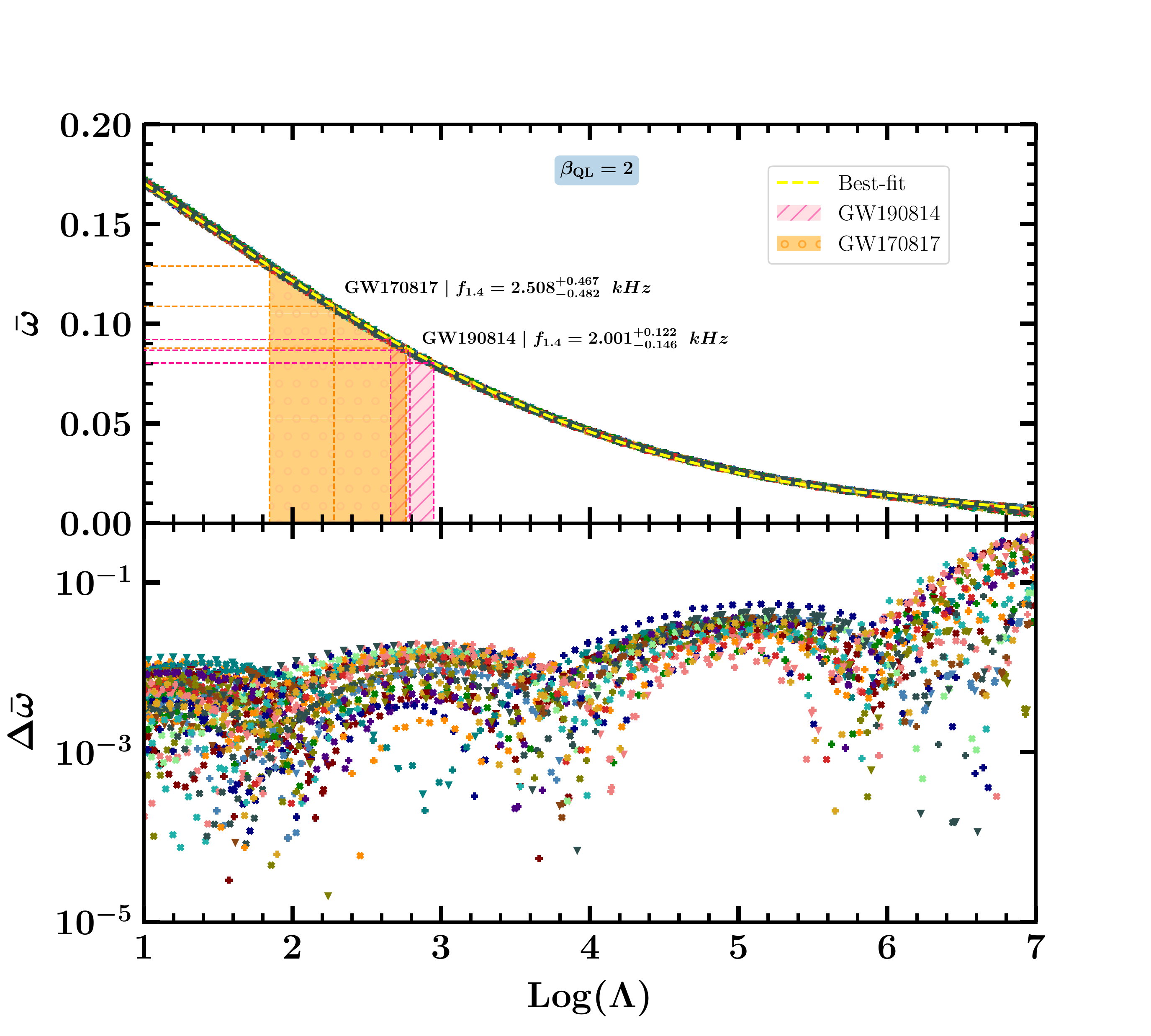}
    \includegraphics[width=0.5\linewidth]{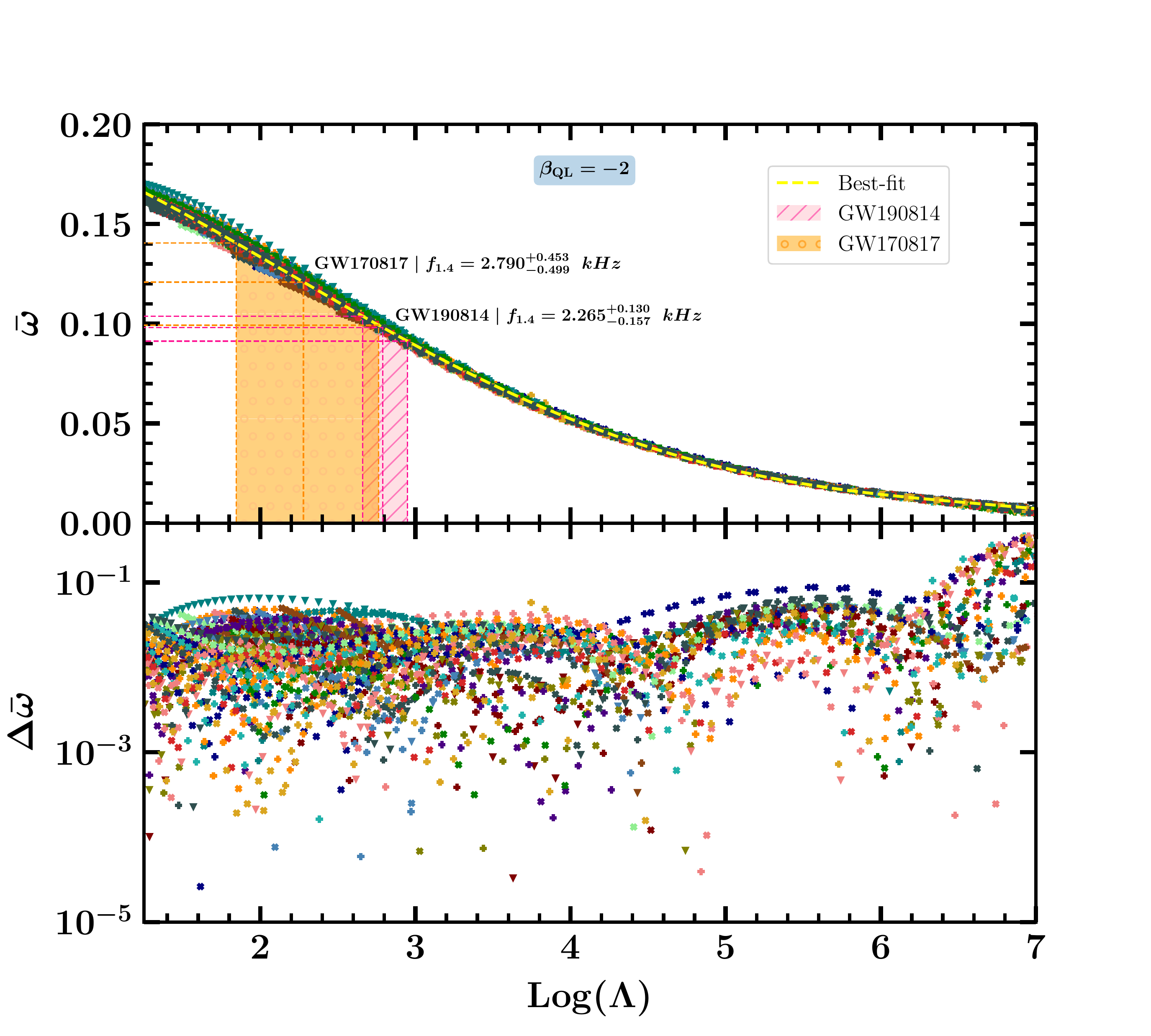}
    \caption{\textit{Left:} Same as Fig. \ref{fig:f-Love | beta= 0}, but with $\beta_{\rm QL}= +2$. \textit{Right:} For $\beta_{\rm QL}= -2$}
    \label{fig:f-Love | beta= -2}
\end{figure}

\subsection{$f$-Love relation}
\label{constraining_f_mode}
An important tool for studying the oscillation of NSs through observational exploration is a UR between the non-radial $f$-mode frequency (a promising source of GWs) and the tidal deformability (a parameter that can be extracted from the GW data). The exploration of multi-polar universal relations between the $f$-mode frequency and tidal deformability of compact stars was first explored by Chan {\it et al.} \cite{Chan_2014} and further improvised by Pradhan {\it et al.} \cite{Bikram_2023}. Recently, Sotani and Kumar \cite{Sotani_2021} introduced a UR between the quasi-normal modes and tidal deformability for isotropic NSs. In this work, we calculate the $f$-Love relations for anisotropic NSs and perform a least-squares fit using the approximate formula
\begin{equation}
\label{eq:f-Love_fitting}
    \Bar{\omega} = \sum_{n=0}^{n=4} d_n (\log(\Lambda))^n \, .
\end{equation}

The coefficients ($d_n$) with $\chi_r^2$ errors are listed in Table \ref{tab:f-Love_coef}. For positive values of anisotropy, the errors in $\chi_r^2$ decrease, indicating that the EOS-insensitive relations become stronger with the addition of anisotropy. Conversely, for negative values, the errors increase. Therefore, positive values of anisotropy strengthen the $f$-Love UR.

\begin{table}
    \begin{minipage}{.48\textwidth}
    \centering
    \caption{The fitting coefficients are listed for $f$-Love relation with $-2.0<\beta_{\mathrm{QL}}<+2.0$. The reduced chi-squared ($\chi_r^2$) is also given for all cases.}
    \renewcommand{\arraystretch}{1.8}
    \scalebox{0.7}{
    \begin{tabular}{cccccc}
        \hline \hline
        $\beta_{\mathrm{QL}}=$ & -2.0 & -1.0 & 0.0 & 1.0 & +2.0 \\
        \hline
        
        $d_0\left(10^{-1}\right)=$ & 2.001 & 2.037 & 2.077 & 2.131 & 2.169  \\
        
        $d_1\left(10^{-2}\right)=$ & -0.964 & -1.998 & -2.722 & -3.607 & -4.158 \\
        
        $d_2\left(10^{-2}\right)=$ & -1.857 & -1.443 & -1.215 & -0.874 & -0.699  \\
        
        $d_3\left(10^{-3}\right)=$ & 3.795 & 3.207 & 2.975 & 2.406 & 2.260 \\
        
        $d_4\left(10^{-4}\right)=$ & -2.155 & -1.873 & -1.815 & -1.538 & -1.464 \\
        
        $\chi_r^2\left(10^{-6}\right)=$ & 4.149 & 2.311 & 1.269 & 0.917 & 0.735  \\ \hline  \hline
    \end{tabular}}
    \label{tab:f-Love_coef}
    \end{minipage}%
    \hfill
    \begin{minipage}{.48\textwidth}
    \centering
    \caption{The canonical normalized $f$-mode frequency $(\Bar{\omega}_{1.4})$, and f-mode frequency ($f_{1.4}$ in kHz) inferred from GW170817 and GW190814 data.}
    \renewcommand{\arraystretch}{1.8}
    \scalebox{0.7}{
    \begin{tabular}{cccccc}
        \hline \hline & \multicolumn{2}{c}{$\mathrm{GW170817}$} & & \multicolumn{2}{c}{$\mathrm{GW190814}$} \\
        \cline { 2 - 3 } \cline { 5 - 6 } $\beta_{\rm QL}$ & $\bar{\omega}_{1.4}$ & $f_{1.4}$ & & $\bar{\omega}_{1.4}$ & $f_{1.4}$ \\
        \hline 
        -2.0 & $0.121^{+0.020} _ {-0.022}$ & $2.790^{+0.453} _ {-0.499}$ & & $0.098^{+0.006} _ {-0.007}$ & $2.265^{+0.130} _ {-0.157}$ \\
    
        -1.0 & $0.116^{+0.020} _ {-0.021}$ & $2.680^{+0.451} _ {-0.487}$ & & $0.094^{+0.005} _ {-0.007}$ & $2.168^{+0.126} _ {-0.152}$ \\
        
        0.0 & $0.113^{+0.020} _ {-0.021}$ & $2.606^{+0.457} _ {-0.484}$ & & $0.091^{+0.005} _ {-0.006}$ & $2.097^{+0.124} _ {-0.149}$ \\
    
        +1.0 & $0.111^{+0.020} _ {-0.021}$ & $2.550^{+0.461} _ {-0.481}$ & & $0.089^{+0.005} _ {-0.006}$ & $2.044^{+0.123} _ {-0.147}$ \\
    
        +2.0 & $0.109^{+0.020} _ {-0.021}$ & $2.508^{+0.467} _ {-0.482}$ & & $0.087^{+0.005} _ {-0.006}$ & $2.001^{+0.122} _ {-0.146}$ \\
        \hline \hline
    \end{tabular}}
    \label{tab:canonical_f-mode}
    \end{minipage} 
\end{table}
\begin{table}
\centering
\caption{The canonical $f$-mode frequency ($f_{1.4}$ in kHz) inferred from GW170817 and GW190814 data using $f$-Love UR obtained in different literature for isotropic NS.}
\renewcommand{\arraystretch}{1.3}
\scalebox{0.9}{
\begin{tabular}{cccc}
    \hline \hline & \multicolumn{1}{c}{$\mathrm{GW170817}$} & & \multicolumn{1}{c}{$\mathrm{GW190814}$} \\
    \cline { 2-2 } \cline { 4-4} Ref. & $f_{1.4}$ & & $f_{1.4}$ \\ 
    \hline \vspace{-3mm} \\
    \parbox[c]{0.35\linewidth}{\centering Chan {\it et al.}  \\ \cite{Chan_2014}} & $2.120 ^{+0.445} _ {-0.446}$ & & $1.652 ^{+0.111} _ {-0.130}$ \vspace{1.5mm} \\

    \parbox[c]{0.35\linewidth}{\centering Pradhan  {\it et al.} \\  \cite{Bikram_2023}} & $2.120 ^{+0.444} _ {-0.445}$ & & $1.653 ^{+0.111} _ {-0.130}$ \vspace{1.5mm} \\
    
    \parbox[c]{0.35\linewidth}{\centering Sotani and Kumar  \\ \cite{Sotani_2021}} & $2.124 ^{+0.440} _ {-0.446}$ & & $1.656 ^{+0.112} _ {-0.132}$ \vspace{1.5mm} \\

     This Work  & $2.606^{+0.457} _ {-0.484}$ & & $2.097^{+0.124} _ {-0.149}$ \vspace{2mm} \\ 
    \hline \hline
\end{tabular}}
\label{tab:comparison}
\end{table}
\subsection{Comparison Study}

We constrain the canonical $f$-mode frequency for GW170817 \cite{GW170817} and GW190814 \cite{GW190814} events across different degrees of anisotropy, as outlined in Table \ref{tab:canonical_f-mode}. The canonical $f$-mode frequency is also compared with previous studies, focusing on isotropic NS, as listed in Table \ref{tab:comparison}. Notably, the $f$-mode frequency obtained in this study is approximately 30-35\% more than the findings of Chan {\it et al.} \cite{Chan_2014}, Pradhan {\it et al.} \cite{Bikram_2023}, and Sotani and Kumar \cite{Sotani_2021}. This difference in the f-mode was anticipated, given that the aforementioned authors employed a full-GR formalism for their $f$-mode calculations, in contrast to our use of the Cowling approximation in this study.

\section{Conclusion}

In this study, we have explored the properties of anisotropic NS with the help of the QL-model proposed by Horvat {\it et al.} \cite{Horvat_2011}. The main motivation for taking the QL-model is that it ensures that $r\rightarrow0$, the anisotropy must vanish, and in other parts of the star, the anisotropy must be there. Different fluid conditions are also studied for varieties of EOSs, and it found that all conditions are perfectly satisfied for the QL model. The speed of sound is also non-negative with any degree of anisotropicity for the QL-model in comparison to the BL-model, as mentioned in Refs. \cite{Bhaskar_Biswas, HC_I_LOVE_C}. Therefore, one can vary the limit of the QL-model from negative to positive values to calculate various properties of the NS. 

Different macroscopic properties of the star have been calculated with different degrees of anisotropy with the help of a variety EOSs spanning from relativistic to non-relativistic cases. It has been observed that the magnitude of the macroscopic properties increases (decreases) for positive (negative) values of $\beta_{\rm QL}$. Almost all the considered EOSs satisfy the different observational limits provided by different observations such as X-ray, pulsar, NICER, GWs, etc. One can impose strong constraints on them with the help of these observational data. Furthermore, we found that positive and negative anisotropy affects tidal deformability parameters and quadrupolar non-radial $f$-mode frequency significantly, which suggests that the star with higher anisotropy sustains more life in the inspiral-merger phase, while the star with lower anisotropy is more likely to collapse.

In addition, we have studied the $I-f-C$ UR for anisotropic NSs for five values of $\beta_{\rm QL}= -2.0, -1.0, 0.0, +1.0$, and $+2.0$. This analysis considered almost 60 tabulated EOS-ensembles spanning a wide range of stiffness, complying with multimessenger constraints. Moreover, one can use the $I-f-C$ universal relation for anisotropic stars to extract information about different properties that are not directly observable with current detectors and telescopes. By varying the anisotropy value, we calculated the $I$-$f$, $C$-$f$, and $C$-$I$ universal relations and fitted them with the polynomial equation using the least-square method. Our results showed that the reduced chi-square errors for the $I-f, C-f,$ and $C$-$I$ relations were $5.4863 \times 10^{-6}, 2.0599 \times 10^{-6}$, and $6.8402 \times 10^{-6}$, respectively, for isotropic stars. In addition to the $I-f-C$ universal relations, we calculated the $f$-Love universal relation to constrain the canonical $f$-mode frequency for anisotropic stars. We observed that the sensitivity of the $C$-$f$ universal relation is weaker for anisotropic stars in comparison to the isotropic case. However, the relation between $I$-$f$ and $f$-Love became stronger with increasing anisotropy. The $C$-$I$ relation barely changed with the inclusion of anisotropy compared to the other universal relations. \SRM{The distribution of $f$-mode across mass-radius parameter space of NSs as obtained by utilizing the $C$-$f$ relation studied for different anisotropic cases.}

With the help of various observational data for dimensionless tidal deformability, such as GW170817 and GW190814, we established a theoretical constraint on the canonical $f$-mode frequency for both isotropic and anisotropic stars, which is presented in Table \ref{tab:canonical_f-mode}. \SRM{As our main objective in this paper was to analyze variations in $I-f-C$ URs resulting from the inclusion of anisotropy, we adhered to the Cowling approximation formalism for computing the $f$-mode. This choice was necessitated by the absence of a comprehensive and reliable full GR formalism for determining QNM in anisotropic NSs. Consequently, for compensation, we calculated constraints on the canonical $f$-mode frequency for isotropic stars, relying on URs obtained by researchers in Ref. \cite{Chan_2014, Bikram_2023, Sotani_2001}, which followed a full-GR formalism, and summarized the outcomes in Table \ref{tab:comparison}.} This constraint can be refined by incorporating different anisotropy models and considering various phenomena such as magnetic fields, quarks in the core, and dark matter in detail in future work. Therefore, our findings provide avenues for investigating the various mechanisms that generate anisotropy within compact stars and for constraining its degree with observational data.
\section{Acknowledgments}
I would like to thank P. Landry and T. Zhao for their fruitful discussions regarding universal relations and fitting procedures. B.K. acknowledges partial support from the Department of Science and Technology, Government of India, with grant no. CRG/2021/000101.
\bibliographystyle{JHEP}
\bibliography{ifc_JCAP.bib} 

\providecommand{\href}[2]{#2}\begingroup\raggedright\begin{thebibliography}{100}

\bibitem{Abbott_2009}
B.P.~Abbott et~al., \emph{Ligo: the laser interferometer gravitational-wave
  observatory},
  \href{https://doi.org/10.1088/0034-4885/72/7/076901}{\emph{Reports on
  Progress in Physics} {\bfseries 72} (2009) 076901}.

\bibitem{Harry_2010}
G.M.~Harry and (forthe LIGO Scientific~Collaboration), \emph{Advanced ligo: the
  next generation of gravitational wave detectors},
  \href{https://doi.org/10.1088/0264-9381/27/8/084006}{\emph{Classical and
  Quantum Gravity} {\bfseries 27} (2010) 084006}.

\bibitem{Acernese_2006}
F.~Acernese et~al., \emph{The virgo status},
  \href{https://doi.org/10.1088/0264-9381/23/19/S01}{\emph{Classical and
  Quantum Gravity} {\bfseries 23} (2006) S635}.

\bibitem{Accadia_2011}
T.~Accadia et~al., \emph{Calibration and sensitivity of the virgo detector
  during its second science run},
  \href{https://doi.org/10.1088/0264-9381/28/2/025005}{\emph{Classical and
  Quantum Gravity} {\bfseries 28} (2010) 025005}.

\bibitem{Antonucci_2011}
F.~Antonucci et~al., \emph{From laboratory experiments to lisa pathfinder:
  achieving lisa geodesic motion},
  \href{https://doi.org/10.1088/0264-9381/28/9/094002}{\emph{Classical and
  Quantum Gravity} {\bfseries 28} (2011) 094002}.

\bibitem{Punturo_2010}
M.~Punturo et~al., \emph{The third generation of gravitational wave
  observatories and their science reach},
  \href{https://doi.org/10.1088/0264-9381/27/8/084007}{\emph{Classical and
  Quantum Gravity} {\bfseries 27} (2010) 084007}.

\bibitem{galaxies10040090}
E.D.~Hall, \emph{Cosmic explorer: A next-generation ground-based
  gravitational-wave observatory},
  \href{https://doi.org/10.3390/galaxies10040090}{\emph{Galaxies} {\bfseries
  10} (2022) }.

\bibitem{Chandrasekhar_1964}
S.~{Chandrasekhar}, \emph{{The Dynamical Instability of Gaseous Masses
  Approaching the Schwarzschild Limit in General Relativity.}},
  \href{https://doi.org/10.1086/147938}{\emph{apj} {\bfseries 140} (1964) 417}.

\bibitem{Chanmugam_1977}
G.~{Chanmugam}, \emph{{Radial oscillations of zero-temperature white dwarfs and
  neutron stars below nuclear densities.}},
  \href{https://doi.org/10.1086/155627}{\emph{apj} {\bfseries 217} (1977) 799}.

\bibitem{Kokkotas_2001}
K.D.~Kokkotas and J.~Ruoff, \emph{Radial oscillations of relativistic stars},
  \href{https://doi.org/10.1051/0004-6361:20000216}{\emph{Astronomy \&
  Astrophysics} {\bfseries 366} (2001) 565}.

\bibitem{Pinku-PRD_2023}
P.~Routaray, H.C.~Das, S.~Sen, B.~Kumar, G.~Panotopoulos and T.~Zhao,
  \emph{Radial oscillations of dark matter admixed neutron stars},
  \href{https://doi.org/10.1103/PhysRevD.107.103039}{\emph{Phys. Rev. D}
  {\bfseries 107} (2023) 103039}.

\bibitem{sen2022radial}
S.~Sen, S.~Kumar, A.~Kunjipurayil, P.~Routaray, S.~Ghosh, P.J.~Kalita et~al.,
  \emph{Radial oscillations in neutron stars from unified hadronic and
  quarkyonic equation of states},
  \href{https://doi.org/10.3390/galaxies11020060}{\emph{Galaxies} {\bfseries
  11} (2023) }.

\bibitem{Pinku_mnras_2023}
P.~Routaray, A.~Quddus, K.~Chakravarti and B.~Kumar, \emph{{Probing the impact
  of WIMP dark matter on universal relations, GW170817 posterior, and radial
  oscillations}}, \href{https://doi.org/10.1093/mnras/stad2628}{\emph{Monthly
  Notices of the Royal Astronomical Society} {\bfseries 525} (2023) 5492}
  [\href{https://arxiv.org/abs/https://academic.oup.com/mnras/article-pdf/525/4/5492/51554155/stad2628.pdf}{{\ttfamily
  https://academic.oup.com/mnras/article-pdf/525/4/5492/51554155/stad2628.pdf}}].

\bibitem{mcdermott_1988}
P.N.~{McDermott}, H.M.~{van Horn} and C.J.~{Hansen}, \emph{{Nonradial
  Oscillations of Neutron Stars}},
  \href{https://doi.org/10.1086/166044}{\emph{\apj} {\bfseries 325} (1988)
  725}.

\bibitem{Kunjipurayil_2022}
A.~Kunjipurayil, T.~Zhao, B.~Kumar, B.K.~Agrawal and M.~Prakash, \emph{Impact
  of the equation of state on $f$- and $p$- mode oscillations of neutron
  stars}, \href{https://doi.org/10.1103/PhysRevD.106.063005}{\emph{Phys. Rev.
  D} {\bfseries 106} (2022) 063005}.

\bibitem{Harish-fmode_2022}
H.C.~Das, A.~Kumar, S.K.~Biswal and S.K.~Patra, \emph{Impacts of dark matter on
  the $f$-mode oscillation of hyperon star},
  \href{https://doi.org/10.1103/PhysRevD.104.123006}{\emph{Phys. Rev. D}
  {\bfseries 104} (2021) 123006}.

\bibitem{Pinku_jcap_2023}
P.~Routaray, S.R.~Mohanty, H.~Das, S.~Ghosh, P.~Kalita, V.~Parmar et~al.,
  \emph{Investigating dark matter-admixed neutron stars with nitr equation of
  state in light of psr j0952-0607},
  \href{https://doi.org/10.1088/1475-7516/2023/10/073}{\emph{Journal of
  Cosmology and Astroparticle Physics} {\bfseries 2023} (2023) 073}.

\bibitem{Zhao_2022}
T.~Zhao and J.M.~Lattimer, \emph{Universal relations for neutron star $f$-mode
  and $g$-mode oscillations},
  \href{https://doi.org/10.1103/PhysRevD.106.123002}{\emph{Phys. Rev. D}
  {\bfseries 106} (2022) 123002}.

\bibitem{PhysRevC.103.035810}
B.K.~Pradhan and D.~Chatterjee, \emph{Effect of hyperons on $f$-mode
  oscillations in neutron stars},
  \href{https://doi.org/10.1103/PhysRevC.103.035810}{\emph{Phys. Rev. C}
  {\bfseries 103} (2021) 035810}.

\bibitem{Pradhan_2022}
B.K.~Pradhan, D.~Chatterjee, M.~Lanoye and P.~Jaikumar, \emph{General
  relativistic treatment of $f$-mode oscillations of hyperonic stars},
  \href{https://doi.org/10.1103/PhysRevC.106.015805}{\emph{Phys. Rev. C}
  {\bfseries 106} (2022) 015805}.

\bibitem{PhysRevD.104.123002}
H.~Sotani and B.~Kumar, \emph{Universal relations between the quasinormal modes
  of neutron star and tidal deformability},
  \href{https://doi.org/10.1103/PhysRevD.104.123002}{\emph{Phys. Rev. D}
  {\bfseries 104} (2021) 123002}.

\bibitem{Finn_1987}
L.S.~Finn, \emph{{g-modes in zero-temperature neutron stars}},
  \href{https://doi.org/10.1093/mnras/227.2.265}{\emph{Mon. Not. R. Astron.
  Soc.} {\bfseries 227} (1987) 265}.

\bibitem{Reisenegger_1992}
A.~{Reisenegger} and P.~{Goldreich}, \emph{{A New Class of g-Modes in Neutron
  Stars}}, \href{https://doi.org/10.1086/171645}{\emph{\apj} {\bfseries 395}
  (1992) 240}.

\bibitem{Zhao_gmode_2022}
T.~Zhao, C.~Constantinou, P.~Jaikumar and M.~Prakash, \emph{Quasinormal $g$
  modes of neutron stars with quarks},
  \href{https://doi.org/10.1103/PhysRevD.105.103025}{\emph{Phys. Rev. D}
  {\bfseries 105} (2022) 103025}.

\bibitem{Lozano_2022}
N.~Lozano, V.~Tran and P.~Jaikumar, \emph{{Temperature Effects on Core g-Modes
  of Neutron Stars}},
  \href{https://doi.org/10.3390/galaxies10040079}{\emph{Galaxies} {\bfseries
  10} (2022) 79}.

\bibitem{Constantinou_2021}
C.~Constantinou, S.~Han, P.~Jaikumar and M.~Prakash, \emph{{g modes of neutron
  stars with hadron-to-quark crossover transitions}},
  \href{https://doi.org/10.1103/physrevd.104.123032}{\emph{Phys. Rev. D}
  {\bfseries 104} (2021) 123032}.

\bibitem{Jaikumar_2021}
P.~Jaikumar, A.~Semposki, M.~Prakash and C.~Constantinou, \emph{{g -mode
  oscillations in hybrid stars: A tale of two sounds}},
  \href{https://doi.org/10.1103/physrevd.103.123009}{\emph{Phys. Rev. D}
  {\bfseries 103} (2021) 123009}.

\bibitem{Wei_2020}
W.~Wei, M.~Salinas, T.~Kl{\ifmmode\ddot{a}\else\"{a}\fi}hn, P.~Jaikumar and
  M.~Barry, \emph{{Lifting the Veil on Quark Matter in Compact Stars with Core
  g-mode Oscillations}},
  \href{https://doi.org/10.3847/1538-4357/abbe02}{\emph{Astrophys. J.}
  {\bfseries 904} (2020) 187}.

\bibitem{Tran_2022}
V.~Tran, S.~Ghosh, N.~Lozano, D.~Chatterjee and P.~Jaikumar, \emph{{g-mode
  Oscillations in Neutron Stars with Hyperons}},  12, 2022.

\bibitem{Haskell_2014}
B.~Haskell, K.~Glampedakis and N.~Andersson, \emph{{A new mechanism for
  saturating unstable r modes in neutron stars}},
  \href{https://doi.org/10.1093/mnras/stu535}{\emph{Mon. Not. R. Astron. Soc.}
  {\bfseries 441} (2014) 1662}.

\bibitem{Haskell_2015}
B.~Haskell, \emph{R-modes in neutron stars: Theory and observations},
  \href{https://doi.org/10.1142/s0218301315410074}{\emph{International Journal
  of Modern Physics E} {\bfseries 24} (2015) 1541007}.

\bibitem{Jyothilakshmi_2022}
O.P.~Jyothilakshmi, P.E.S.~Krishnan, P.~Thakur, V.~Sreekanth and T.K.~Jha,
  \emph{{Hyperon bulk viscosity and r-modes of neutron stars}},
  \href{https://doi.org/10.1093/mnras/stac2360}{\emph{Mon. Not. R. Astron.
  Soc.} {\bfseries 516} (2022) 3381}.

\bibitem{Lin_2004}
L.-M.~{Lin}, \emph{{Numerical study of nonlinear R-modes in neutron stars}},
  Ph.D. thesis, Washington University in Saint Louis, Missouri, Jan., 2004.

\bibitem{Rezzolla_2002}
L.~Rezzolla, \emph{{The r-modes Oscillations and Instability: Surprises from
  Magnetized Neutron Stars}},  in \emph{{Recent Developments in General
  Relativity}}, pp.~235--248, Springer, Milano (2002),
  \href{https://doi.org/10.1007/978-88-470-2101-3_16}{DOI}.

\bibitem{Michael_2017}
M.~Jasiulek and C.~Chirenti, \emph{$r$-mode frequencies of rapidly and
  differentially rotating relativistic neutron stars},
  \href{https://doi.org/10.1103/PhysRevD.95.064060}{\emph{Phys. Rev. D}
  {\bfseries 95} (2017) 064060}.

\bibitem{Benhar_1999}
O.~Benhar, E.~Berti and V.~Ferrari, \emph{{The imprint of the equation of state
  on the axial w-modes of oscillating neutron stars}},
  \href{https://doi.org/10.1046/j.1365-8711.1999.02983.x}{\emph{Mon. Not. R.
  Astron. Soc.} {\bfseries 310} (1999) 797}.

\bibitem{Bandyopadhyay_2012}
D.~Bandyopadhyay and D.~Chatterjee, \emph{{AXIAL W-MODES OF NEUTRON STARS WITH
  EXOTIC MATTER}},
  \href{https://doi.org/10.1142/9789814374552_0102}{\emph{WORLD SCIENTIFIC}
  (2012) 949}.

\bibitem{Kokkotas1999}
K.D.~Kokkotas and B.G.~Schmidt, \emph{Quasi-normal modes of stars and black
  holes}, \href{https://doi.org/10.12942/lrr-1999-2}{\emph{Living Reviews in
  Relativity} {\bfseries 2} (1999) 2}.

\bibitem{Sotani_2011}
H.~Sotani, N.~Yasutake, T.~Maruyama et~al., \emph{Signatures of hadron-quark
  mixed phase in gravitational waves},
  \href{https://doi.org/10.1103/PhysRevD.83.024014}{\emph{Phys. Rev. D}
  {\bfseries 83} (2011) 024014}.

\bibitem{Flores_2014}
C.V.~Flores and G.~Lugones, \emph{Discriminating hadronic and quark stars
  through gravitational waves of fluid pulsation modes},
  \href{https://doi.org/10.1088/0264-9381/31/15/155002}{\emph{Classical and
  Quantum Gravity} {\bfseries 31} (2014) 155002}.

\bibitem{Sandoval_2018}
I.F.~Ranea-Sandoval, O.M.~Guilera, M.~Mariani et~al., \emph{Oscillation modes
  of hybrid stars within the relativistic cowling approximation},
  \href{https://doi.org/10.1088/1475-7516/2018/12/031}{\emph{Journal of
  Cosmology and Astroparticle Physics} {\bfseries 2018} (2018) 031}.

\bibitem{Tianqi_2022}
T.~Zhao, C.~Constantinou, P.~Jaikumar et~al., \emph{Quasinormal $g$ modes of
  neutron stars with quarks},
  \href{https://doi.org/10.1103/PhysRevD.105.103025}{\emph{Phys. Rev. D}
  {\bfseries 105} (2022) 103025}.

\bibitem{Shibagaki_2020}
S.~Shibagaki, T.~Kuroda, K.~Kotake et~al., \emph{{A new gravitational-wave
  signature of low-T/|W| instability in rapidly rotating stellar core
  collapse}}, \href{https://doi.org/10.1093/mnrasl/slaa021}{\emph{Monthly
  Notices of the Royal Astronomical Society: Letters} {\bfseries 493} (2020)
  L138}.

\bibitem{Andersson_1998}
N.~Andersson and K.D.~Kokkotas, \emph{{Towards gravitational wave
  asteroseismology}},
  \href{https://doi.org/10.1046/j.1365-8711.1998.01840.x}{\emph{Monthly Notices
  of the Royal Astronomical Society} {\bfseries 299} (1998) 1059}.

\bibitem{Lau_2010}
H.K.~Lau, P.T.~Leung and L.M.~Lin, \emph{Inferring physical parameters of
  compact stars from their f-mode gravitational wave signals},
  \href{https://doi.org/10.1088/0004-637X/714/2/1234}{\emph{The Astrophysical
  Journal} {\bfseries 714} (2010) 1234}.

\bibitem{Chan_2014}
T.K.~Chan, Y.-H.~Sham, P.T.~Leung et~al., \emph{Multipolar universal relations
  between $f$-mode frequency and tidal deformability of compact stars},
  \href{https://doi.org/10.1103/PhysRevD.90.124023}{\emph{Phys. Rev. D}
  {\bfseries 90} (2014) 124023}.

\bibitem{Sotani_2021}
H.~Sotani and B.~Kumar, \emph{Universal relations between the quasinormal modes
  of neutron star and tidal deformability},
  \href{https://doi.org/10.1103/PhysRevD.104.123002}{\emph{Phys. Rev. D}
  {\bfseries 104} (2021) 123002}.

\bibitem{Bikram_2023}
B.K.~Pradhan, A.~Vijaykumar and D.~Chatterjee, \emph{Impact of updated
  multipole love numbers and $f$-love universal relations in the context of
  binary neutron stars},
  \href{https://doi.org/10.1103/PhysRevD.107.023010}{\emph{Phys. Rev. D}
  {\bfseries 107} (2023) 023010}.

\bibitem{Kent_yagi_2013}
K.~Yagi and N.~Yunes, \emph{I-love-q relations in neutron stars and their
  applications to astrophysics, gravitational waves, and fundamental physics},
  \href{https://doi.org/10.1103/PhysRevD.88.023009}{\emph{Phys. Rev. D}
  {\bfseries 88} (2013) 023009}.

\bibitem{Kent_yagi_2015}
K.~Yagi and N.~Yunes, \emph{I-love-q anisotropically: Universal relations for
  compact stars with scalar pressure anisotropy},
  \href{https://doi.org/10.1103/PhysRevD.91.123008}{\emph{Phys. Rev. D}
  {\bfseries 91} (2015) 123008}.

\bibitem{Breu_2016}
C.~Breu and L.~Rezzolla, \emph{{Maximum mass, moment of inertia and compactness
  of relativistic stars}},
  \href{https://doi.org/10.1093/mnras/stw575}{\emph{Monthly Notices of the
  Royal Astronomical Society} {\bfseries 459} (2016) 646}.

\bibitem{Rezzolla_2018}
L.~Rezzolla, P.~Pizzochero, D.~Jones, N.~Rea and I.~Vida{\~n}a, \emph{The
  Physics and Astrophysics of Neutron Stars}, Astrophysics and Space Science
  Library, Springer International Publishing (2019).

\bibitem{Riahi_2019}
R.~Riahi, S.Z.~Kalantari and J.A.~Rueda, \emph{Universal relations for the
  keplerian sequence of rotating neutron stars},
  \href{https://doi.org/10.1103/PhysRevD.99.043004}{\emph{Phys. Rev. D}
  {\bfseries 99} (2019) 043004}.

\bibitem{Gupta_2018}
T.~Gupta, B.~Majumder, K.~Yagi et~al., \emph{I-love-q relations for neutron
  stars in dynamical chern simons gravity},
  \href{https://doi.org/10.1088/1361-6382/aa9c68}{\emph{Classical and Quantum
  Gravity} {\bfseries 35} (2017) 025009}.

\bibitem{Jiang_2020}
J.-L.~Jiang, S.-P.~Tang, Y.-Z.~Wang et~al., \emph{Psr j0030+0451, gw170817, and
  the nuclear data: Joint constraints on equation of state and bulk properties
  of neutron stars}, \href{https://doi.org/10.3847/1538-4357/ab77cf}{\emph{The
  Astrophysical Journal} {\bfseries 892} (2020) 55}.

\bibitem{universe7040111}
C.-H.~Yeung, L.-M.~Lin, N.~Andersson et~al., \emph{The i-love-q relations for
  superfluid neutron stars},
  \href{https://doi.org/10.3390/universe7040111}{\emph{Universe} {\bfseries 7}
  (2021) }.

\bibitem{Chakrabarti_2014}
S.~Chakrabarti, T.~Delsate, N.~G\"urlebeck et~al.,
  \emph{$i\text{\ensuremath{-}}q$ relation for rapidly rotating neutron stars},
  \href{https://doi.org/10.1103/PhysRevLett.112.201102}{\emph{Phys. Rev. Lett.}
  {\bfseries 112} (2014) 201102}.

\bibitem{Haskell_2013}
B.~Haskell, R.~Ciolfi, F.~Pannarale et~al., \emph{{On the universality of
  I–Love–Q relations in magnetized neutron stars}},
  \href{https://doi.org/10.1093/mnrasl/slt161}{\emph{Monthly Notices of the
  Royal Astronomical Society: Letters} {\bfseries 438} (2013) L71}.

\bibitem{Bandyopadhyay2018}
D.~Bandyopadhyay, S.A.~Bhat, P.~Char et~al., \emph{Moment of inertia,
  quadrupole moment, love number of neutron star and their relations with
  strange-matter equations of state},
  \href{https://doi.org/10.1140/epja/i2018-12456-y}{\emph{The European Physical
  Journal A} {\bfseries 54} (2018) 26}.

\bibitem{S.S.Yaza}
S.S.~Yazadjiev, \emph{Relativistic models of magnetars: Nonperturbative
  analytical approach},
  \href{https://doi.org/10.1103/PhysRevD.85.044030}{\emph{Phys. Rev. D}
  {\bfseries 85} (2012) 044030}.

\bibitem{C.Y.Cardall_2001}
C.Y.~Cardall, M.~Prakash and J.M.~Lattimer, \emph{Effects of strong magnetic
  fields on neutron star structure},
  \href{https://doi.org/10.1086/321370}{\emph{The Astrophysical Journal}
  {\bfseries 554} (2001) 322}.

\bibitem{K.Ioka_2004}
K.~Ioka and M.~Sasaki, \emph{Relativistic stars with poloidal and toroidal
  magnetic fields and meridional flow},
  \href{https://doi.org/10.1086/379650}{\emph{The Astrophysical Journal}
  {\bfseries 600} (2004) 296}.

\bibitem{R.Ciolfi_L.Rezzolla}
R.~Ciolfi and L.~Rezzolla, \emph{{Twisted-torus configurations with large
  toroidal magnetic fields in relativistic stars}},
  \href{https://doi.org/10.1093/mnrasl/slt092}{\emph{Monthly Notices of the
  Royal Astronomical Society: Letters} {\bfseries 435} (2013) L43}.

\bibitem{R.Ciolfi_V.Ferrari}
R.~Ciolfi, V.~Ferrari and L.~Gualtieri, \emph{{Structure and deformations of
  strongly magnetized neutron stars with twisted-torus configurations}},
  \href{https://doi.org/10.1111/j.1365-2966.2010.16847.x}{\emph{Monthly Notices
  of the Royal Astronomical Society} {\bfseries 406} (2010) 2540}.

\bibitem{Frieben}
J.~Frieben and L.~Rezzolla, \emph{{Equilibrium models of relativistic stars
  with a toroidal magnetic field}},
  \href{https://doi.org/10.1111/j.1365-2966.2012.22027.x}{\emph{Monthly Notices
  of the Royal Astronomical Society} {\bfseries 427} (2012) 3406}.

\bibitem{A.G.Pili}
A.G.~Pili, N.~Bucciantini and L.~Del~Zanna, \emph{{Axisymmetric equilibrium
  models for magnetized neutron stars in General Relativity under the
  Conformally Flat Condition}},
  \href{https://doi.org/10.1093/mnras/stu215}{\emph{Monthly Notices of the
  Royal Astronomical Society} {\bfseries 439} (2014) 3541}.

\bibitem{N.Bucciantini}
N.~Bucciantini, A.G.~Pili and L.~Del~Zanna, \emph{{The role of currents
  distribution in general relativistic equilibria of magnetized neutron
  stars}}, \href{https://doi.org/10.1093/mnras/stu2689}{\emph{Monthly Notices
  of the Royal Astronomical Society} {\bfseries 447} (2015) 3278}.

\bibitem{R.F.Sawyer}
R.F.~Sawyer, \emph{Condensed ${\ensuremath{\pi}}^{\ensuremath{-}}$ phase in
  neutron-star matter},
  \href{https://doi.org/10.1103/PhysRevLett.29.382}{\emph{Phys. Rev. Lett.}
  {\bfseries 29} (1972) 382}.

\bibitem{B.Carter}
B.~Carter and D.~Langlois, \emph{Relativistic models for
  superconducting-superfluid mixtures},
  \href{https://doi.org/https://doi.org/10.1016/S0550-3213(98)00430-1}{\emph{Nuclear
  Physics B} {\bfseries 531} (1998) 478}.

\bibitem{V.Canuto}
V.~Canuto, \emph{Equation of state at ultrahigh densities},
  \href{https://doi.org/10.1146/annurev.aa.12.090174.001123}{\emph{Annual
  Review of Astronomy and Astrophysics} {\bfseries 12} (1974) 167}.

\bibitem{M.Ruderman}
M.~Ruderman, \emph{Pulsars: Structure and dynamics},
  \href{https://doi.org/10.1146/annurev.aa.10.090172.002235}{\emph{Annual
  Review of Astronomy and Astrophysics} {\bfseries 10} (1972) 427}.

\bibitem{S.Nelmes}
S.~Nelmes and B.M.A.G.~Piette, \emph{Phase transition and anisotropic
  deformations of neutron star matter},
  \href{https://doi.org/10.1103/PhysRevD.85.123004}{\emph{Phys. Rev. D}
  {\bfseries 85} (2012) 123004}.

\bibitem{W.A.Kippenhahn_Rudolf}
R.~{Kippenhahn} and A.~{Weigert}, \emph{{Stellar Structure and Evolution}},
  1990.

\bibitem{Glendenning:1997wn}
N.K.~Glendenning, \emph{{Compact stars}},  1997.

\bibitem{H.Heiselberg}
H.~Heiselberg and M.~Hjorth-Jensen, \emph{Phases of dense matter in neutron
  stars},
  \href{https://doi.org/https://doi.org/10.1016/S0370-1573(99)00110-6}{\emph{Physics
  Reports} {\bfseries 328} (2000) 237}.

\bibitem{BL-Model}
R.L.~{Bowers} and E.P.T.~{Liang}, \emph{{Anisotropic Spheres in General
  Relativity}}, \href{https://doi.org/10.1086/152760}{\emph{\apj} {\bfseries
  188} (1974) 657}.

\bibitem{Horvat_2011}
D.~Horvat, S.~Ilijić and A.~Marunović, \emph{Radial pulsations and stability
  of anisotropic stars with a quasi-local equation of state},
  \href{https://doi.org/10.1088/0264-9381/28/2/025009}{\emph{Classical and
  Quantum Gravity} {\bfseries 28} (2010) 025009}.

\bibitem{Cosenza}
M.~Cosenza, L.~Herrera, M.~Esculpi et~al., \emph{Some models of anisotropic
  spheres in general relativity},
  \href{https://doi.org/10.1063/1.524742}{\emph{Journal of Mathematical
  Physics} {\bfseries 22} (1981) 118}.

\bibitem{Silva_2015}
H.O.~Silva, C.F.B.~Macedo, E.~Berti et~al., \emph{Slowly rotating anisotropic
  neutron stars in general relativity and scalar–tensor theory},
  \href{https://doi.org/10.1088/0264-9381/32/14/145008}{\emph{Classical and
  Quantum Gravity} {\bfseries 32} (2015) 145008}.

\bibitem{Hillebrandt}
W.~{Hillebrandt} and K.O.~{Steinmetz}, \emph{{Anisotropic neutron star models:
  stability against radial and nonradial pulsations.}}, {\emph{aap} {\bfseries
  53} (1976) 283}.

\bibitem{Doneva_2012}
D.D.~Doneva and S.S.~Yazadjiev, \emph{Nonradial oscillations of anisotropic
  neutron stars in the cowling approximation},
  \href{https://doi.org/10.1103/PhysRevD.85.124023}{\emph{Phys. Rev. D}
  {\bfseries 85} (2012) 124023}.

\bibitem{Bayin}
S.S.~Bayin, \emph{Anisotropic fluid spheres in general relativity},
  \href{https://doi.org/10.1103/PhysRevD.26.1262}{\emph{Phys. Rev. D}
  {\bfseries 26} (1982) 1262}.

\bibitem{Roupas2021}
Z.~Roupas, \emph{Secondary component of gravitational-wave signal gw190814 as
  an anisotropic neutron star},
  \href{https://doi.org/10.1007/s10509-021-03919-5}{\emph{Astrophysics and
  Space Science} {\bfseries 366} (2021) 9}.

\bibitem{Deb_2021}
D.~Deb, B.~Mukhopadhyay and F.~Weber, \emph{Effects of anisotropy on strongly
  magnetized neutron and strange quark stars in general relativity},
  \href{https://doi.org/10.3847/1538-4357/ac222a}{\emph{The Astrophysical
  Journal} {\bfseries 922} (2021) 149}.

\bibitem{Estevez-Delgado2018}
G.~Estevez-Delgado and J.~Estevez-Delgado, \emph{On the effect of anisotropy on
  stellar models},
  \href{https://doi.org/10.1140/epjc/s10052-018-6151-z}{\emph{The European
  Physical Journal C} {\bfseries 78} (2018) 673}.

\bibitem{Pattersons2021}
M.L.~Pattersons and A.~Sulaksono, \emph{Mass correction and deformation of
  slowly rotating anisotropic neutron stars based on hartle--thorne formalism},
  \href{https://doi.org/10.1140/epjc/s10052-021-09481-2}{\emph{The European
  Physical Journal C} {\bfseries 81} (2021) 698}.

\bibitem{Rizaldy}
R.~Rizaldy, A.R.~Alfarasyi, A.~Sulaksono et~al., \emph{Neutron-star deformation
  due to anisotropic momentum distribution of neutron-star matter},
  \href{https://doi.org/10.1103/PhysRevC.100.055804}{\emph{Phys. Rev. C}
  {\bfseries 100} (2019) 055804}.

\bibitem{Rahmansyah2020}
A.~Rahmansyah, A.~Sulaksono, A.B.~Wahidin et~al., \emph{Anisotropic neutron
  stars with hyperons: implication of the recent nuclear matter data and
  observations of neutron stars},
  \href{https://doi.org/10.1140/epjc/s10052-020-8361-4}{\emph{The European
  Physical Journal C} {\bfseries 80} (2020) 769}.

\bibitem{Rahmansyah2021}
A.~Rahmansyah and A.~Sulaksono, \emph{Recent multimessenger constraints and the
  anisotropic neutron star},
  \href{https://doi.org/10.1103/PhysRevC.104.065805}{\emph{Phys. Rev. C}
  {\bfseries 104} (2021) 065805}.

\bibitem{Herrera2008}
L.~Herrera, J.~Ospino and A.~Di~Prisco, \emph{All static spherically symmetric
  anisotropic solutions of einstein's equations},
  \href{https://doi.org/10.1103/PhysRevD.77.027502}{\emph{Phys. Rev. D}
  {\bfseries 77} (2008) 027502}.

\bibitem{Herrera2013}
L.~Herrera and W.~Barreto, \emph{General relativistic polytropes for
  anisotropic matter: The general formalism and applications},
  \href{https://doi.org/10.1103/PhysRevD.88.084022}{\emph{Phys. Rev. D}
  {\bfseries 88} (2013) 084022}.

\bibitem{Bhaskar_Biswas}
B.~Biswas and S.~Bose, \emph{Tidal deformability of an anisotropic compact
  star: Implications of gw170817},
  \href{https://doi.org/10.1103/PhysRevD.99.104002}{\emph{Phys. Rev. D}
  {\bfseries 99} (2019) 104002}.

\bibitem{S.Das2021}
S.~Das, B.K.~Parida, S.~Ray et~al., \emph{Role of anisotropy on the tidal
  deformability of compact stellar objects},
  \href{https://doi.org/10.3390/ECU2021-09311}{\emph{Physical Sciences Forum}
  {\bfseries 2} (2021) }.

\bibitem{Roupas2020}
Z.~Roupas and G.G.L.~Nashed, \emph{Anisotropic neutron stars modelling:
  constraints in krori--barua spacetime},
  \href{https://doi.org/10.1140/epjc/s10052-020-08462-1}{\emph{The European
  Physical Journal C} {\bfseries 80} (2020) 905}.

\bibitem{Sulaksono2015}
A.~{Sulaksono}, \emph{{Anisotropic pressure and hyperons in neutron stars}},
  \href{https://doi.org/10.1142/S021830131550007X}{\emph{International Journal
  of Modern Physics E} {\bfseries 24} (2015) 1550007}.

\bibitem{Setiawan2019}
A.M.~Setiawan and A.~Sulaksono, \emph{Anisotropic neutron stars and perfect
  fluid's energy conditions},
  \href{https://doi.org/10.1140/epjc/s10052-019-7265-7}{\emph{The European
  Physical Journal C} {\bfseries 79} (2019) 755}.

\bibitem{HC_I_LOVE_C}
H.C.~Das, \emph{$i\ensuremath{-}\mathrm{Love}\ensuremath{-}c$ relation for an
  anisotropic neutron star},
  \href{https://doi.org/10.1103/PhysRevD.106.103518}{\emph{Phys. Rev. D}
  {\bfseries 106} (2022) 103518}.

\bibitem{Fortin_2016}
M.~Fortin, C.~Provid\^encia, A.R.~Raduta et~al., \emph{Neutron star radii and
  crusts: Uncertainties and unified equations of state},
  \href{https://doi.org/10.1103/PhysRevC.94.035804}{\emph{Phys. Rev. C}
  {\bfseries 94} (2016) 035804}.

\bibitem{Bharat_and_Landry}
B.~Kumar and P.~Landry, \emph{Inferring neutron star properties from gw170817
  with universal relations},
  \href{https://doi.org/10.1103/PhysRevD.99.123026}{\emph{Phys. Rev. D}
  {\bfseries 99} (2019) 123026}.

\bibitem{Landry_2018}
P.~Landry and B.~Kumar, \emph{Constraints on the moment of inertia of psr
  j0737-3039a from gw170817},
  \href{https://doi.org/10.3847/2041-8213/aaee76}{\emph{The Astrophysical
  Journal Letters} {\bfseries 868} (2018) L22}.

\bibitem{Athul_2022}
A.~Kunjipurayil, T.~Zhao, B.~Kumar et~al., \emph{Impact of the equation of
  state on $f$- and $p$- mode oscillations of neutron stars},
  \href{https://doi.org/10.1103/PhysRevD.106.063005}{\emph{Phys. Rev. D}
  {\bfseries 106} (2022) 063005}.

\bibitem{Tuhin_2018}
T.~Malik, N.~Alam, M.~Fortin et~al., \emph{Gw170817: Constraining the nuclear
  matter equation of state from the neutron star tidal deformability},
  \href{https://doi.org/10.1103/PhysRevC.98.035804}{\emph{Phys. Rev. C}
  {\bfseries 98} (2018) 035804}.

\bibitem{Alam_2016}
N.~Alam, B.K.~Agrawal, M.~Fortin et~al., \emph{Strong correlations of neutron
  star radii with the slopes of nuclear matter incompressibility and symmetry
  energy at saturation},
  \href{https://doi.org/10.1103/PhysRevC.94.052801}{\emph{Phys. Rev. C}
  {\bfseries 94} (2016) 052801}.

\bibitem{Vishal_2022_Crustal}
V.~Parmar, H.C.~Das, A.~Kumar et~al., \emph{Crustal properties of a neutron
  star within an effective relativistic mean-field model},
  \href{https://doi.org/10.1103/PhysRevD.105.043017}{\emph{Phys. Rev. D}
  {\bfseries 105} (2022) 043017}.

\bibitem{Vishal_2022_Pasta}
V.~Parmar, H.C.~Das, A.~Kumar et~al., \emph{Pasta properties of the neutron
  star within effective relativistic mean-field model},
  \href{https://doi.org/10.1103/PhysRevD.106.023031}{\emph{Phys. Rev. D}
  {\bfseries 106} (2022) 023031}.

\bibitem{mod_TOV}
P.~Bhar and P.~Rej, \emph{Compact stellar model in the presence of pressure
  anisotropy in modified finch skea space--time},
  \href{https://doi.org/10.1007/s12036-021-09739-x}{\emph{Journal of
  Astrophysics and Astronomy} {\bfseries 42} (2021) 74}.

\bibitem{TOV}
J.R.~Oppenheimer and G.M.~Volkoff, \emph{On massive neutron cores},
  \href{https://doi.org/10.1103/PhysRev.55.374}{\emph{Phys. Rev.} {\bfseries
  55} (1939) 374}.

\bibitem{QL_Model}
D.~Horvat, S.~Ilijić and A.~Marunović, \emph{Radial pulsations and stability
  of anisotropic stars with a quasi-local equation of state},
  \href{https://doi.org/10.1088/0264-9381/28/2/025009}{\emph{Classical and
  Quantum Gravity} {\bfseries 28} (2010) 025009}.

\bibitem{Sulaksono}
A.M.~Setiawan and A.~Sulaksono, \emph{Anisotropic neutron stars and perfect
  fluid's energy conditions},
  \href{https://doi.org/10.1140/epjc/s10052-019-7265-7}{\emph{The European
  Physical Journal C} {\bfseries 79} (2019) 755}.

\bibitem{Miller_2019}
M.C.~Miller et~al., \emph{Psr j0030+0451 mass and radius from nicer data and
  implications for the properties of neutron star matter},
  \href{https://doi.org/10.3847/2041-8213/ab50c5}{\emph{The Astrophysical
  Journal Letters} {\bfseries 887} (2019) L24}.

\bibitem{Riley_2019}
T.E.~Riley et~al., \emph{A nicer view of psr j0030+0451: Millisecond pulsar
  parameter estimation},
  \href{https://doi.org/10.3847/2041-8213/ab481c}{\emph{The Astrophysical
  Journal Letters} {\bfseries 887} (2019) L21}.

\bibitem{Miller_2021}
M.C.~Miller et~al., \emph{The radius of psr j0740+6620 from nicer and
  xmm-newton data}, \href{https://doi.org/10.3847/2041-8213/ac089b}{\emph{The
  Astrophysical Journal Letters} {\bfseries 918} (2021) L28}.

\bibitem{GW170817}
{\scshape LIGO Scientific Collaboration and Virgo Collaboration} collaboration,
  \emph{Gw170817: Observation of gravitational waves from a binary neutron star
  inspiral}, \href{https://doi.org/10.1103/PhysRevLett.119.161101}{\emph{Phys.
  Rev. Lett.} {\bfseries 119} (2017) 161101}.

\bibitem{GW190814}
R.~Abbott, T.D.~Abbott, L.S.~Collaboration et~al., \emph{Gw190814:
  Gravitational waves from the coalescence of a 23 solar mass black hole with a
  2.6 solar mass compact object},
  \href{https://doi.org/10.3847/2041-8213/ab960f}{\emph{The Astrophysical
  Journal Letters} {\bfseries 896} (2020) L44}.

\bibitem{Hinderer_2008}
T.~Hinderer, \emph{Tidal love numbers of neutron stars},
  \href{https://doi.org/10.1086/533487}{\emph{The Astrophysical Journal}
  {\bfseries 677} (2008) 1216}.

\bibitem{Hinderer_2009}
T.~Hinderer, \emph{Erratum: “tidal love numbers of neutron stars” (2008,
  apj, 677, 1216)},
  \href{https://doi.org/10.1088/0004-637X/697/1/964}{\emph{The Astrophysical
  Journal} {\bfseries 697} (2009) 964}.

\bibitem{Cowling_1941}
T.G.~Cowling, \emph{{The Non-radial Oscillations of Polytropic Stars}},
  \href{https://doi.org/10.1093/mnras/101.8.367}{\emph{Monthly Notices of the
  Royal Astronomical Society} {\bfseries 101} (1941) 367}.

\bibitem{Curi2022}
E.J.A.~Curi, L.B.~Castro, C.V.~Flores and C.H.~Lenzi, \emph{Non-radial
  oscillations and global stellar properties of anisotropic compact stars using
  realistic equations of state},
  \href{https://doi.org/10.1140/epjc/s10052-022-10498-4}{\emph{The European
  Physical Journal C} {\bfseries 82} (2022) 527}.

\bibitem{PhysRevD.101.124006}
N.~Jiang and K.~Yagi, \emph{Analytic i-love-c relations for realistic neutron
  stars}, \href{https://doi.org/10.1103/PhysRevD.101.124006}{\emph{Phys. Rev.
  D} {\bfseries 101} (2020) 124006}.

\bibitem{PhysRevD.91.044034}
C.~Chirenti, G.H.~de~Souza and W.~Kastaun, \emph{Fundamental oscillation modes
  of neutron stars: Validity of universal relations},
  \href{https://doi.org/10.1103/PhysRevD.91.044034}{\emph{Phys. Rev. D}
  {\bfseries 91} (2015) 044034}.

\bibitem{Yagi:2013bca}
K.~Yagi and N.~Yunes, \emph{{I-Love-Q}},
  \href{https://doi.org/10.1126/science.1236462}{\emph{Science} {\bfseries 341}
  (2013) 365} [\href{https://arxiv.org/abs/1302.4499}{{\ttfamily 1302.4499}}].

\bibitem{staykov2016}
K.V.~Staykov, D.D.~Doneva and S.S.~Yazadjiev,
  \emph{Moment-of-inertia--compactness universal relations in scalar-tensor
  theories and ${\mathcal{r}}^{2}$ gravity},
  \href{https://doi.org/10.1103/PhysRevD.93.084010}{\emph{Phys. Rev. D}
  {\bfseries 93} (2016) 084010}.

\bibitem{Ravenhall_1994}
D.G.~{Ravenhall} and C.J.~{Pethick}, \emph{{Neutron Star Moments of Inertia}},
  \href{https://doi.org/10.1086/173935}{\emph{\apj} {\bfseries 424} (1994)
  846}.

\bibitem{Lattimer_2005}
J.M.~Lattimer and B.F.~Schutz, \emph{Constraining the equation of state with
  moment of inertia measurements},
  \href{https://doi.org/10.1086/431543}{\emph{The Astrophysical Journal}
  {\bfseries 629} (2005) 979}.

\bibitem{Popchev2019}
D.~Popchev, K.V.~Staykov, D.D.~Doneva et~al., \emph{Moment of inertia--mass
  universal relations for neutron stars in scalar-tensor theory with
  self-interacting massive scalar field},
  \href{https://doi.org/10.1140/epjc/s10052-019-6691-x}{\emph{The European
  Physical Journal C} {\bfseries 79} (2019) 178}.

\bibitem{refId0}
{Bejger, M.} and {Haensel, P.}, \emph{Moments of inertia for neutron and
  strange stars: Limits derived for the crab pulsar},
  \href{https://doi.org/10.1051/0004-6361:20021241}{\emph{A\&A} {\bfseries 396}
  (2002) 917}.

\bibitem{Sotani_2001}
H.~Sotani, K.~Tominaga and K.-i.~Maeda, \emph{Density discontinuity of a
  neutron star and gravitational waves},
  \href{https://doi.org/10.1103/PhysRevD.65.024010}{\emph{Phys. Rev. D}
  {\bfseries 65} (2001) 024010}.

\end{thebibliography}\endgroup
\end{document}